\documentclass[12pt]{iopart}
\usepackage{graphicx}
\usepackage{psfrag}

\psfrag{Ccal}[B1][B1][1]{${\mathcal{C}}/R$}
\psfrag{Ccal1}[B1][B1][1]{${\mathcal{C}}$}
\psfrag{Vcal}[B1][B1][1]{${\mathcal{V}}$}
\psfrag{Ecal}[B1][B1][1]{${\mathcal{E}}$}

\begin{document}

\topical[Forward flux sampling]{Forward Flux Sampling for rare event simulations}


\author{Rosalind J Allen$^1$, Chantal Valeriani$^1$ and Pieter Rein ten Wolde$^2$}
\address{1. SUPA, School of Physics and Astronomy, The University of Edinburgh, 
Mayfield Road, Edinburgh EH9 3JZ, UK.}
\address{2. FOM Institute for Atomic and Molecular Physics, Science Park 113, 1098 XG Amsterdam, The Netherlands}

\ead{rallen2@ph.ed.ac.uk}

\begin{abstract}
Rare events are ubiquitous in many different fields, yet they are notoriously difficult to simulate because few, if any, events are observed in a conventional simulation run. Over the past several decades, specialised simulation methods have been developed to overcome this problem. We review one recently-developed class of such methods, known as Forward Flux Sampling. Forward Flux Sampling uses a series of interfaces between the initial and final states to calculate rate constants and generate transition paths, for rare events in equilibrium or nonequilibrium systems with stochastic dynamics. This review draws together a number of recent advances, summarises several applications of the method and highlights challenges that remain to be overcome. 
\end{abstract}

\maketitle
\newpage

\contentsline {section}{\numberline {1}Introduction}{3}
\contentsline {section}{\numberline {2}Background}{4}
\contentsline {subsection}{\numberline {2.1}``Bennett-Chandler'' type methods}{4}
\contentsline {subsection}{\numberline {2.2}Transition Path Sampling}{5}
\contentsline {subsection}{\numberline {2.3}Transition Interface Sampling}{6}
\contentsline {subsection}{\numberline {2.4}Milestoning}{7}
\contentsline {subsection}{\numberline {2.5}The Weighted Ensemble method}{7}
\contentsline {subsection}{\numberline {2.6}The Finite Temperature String Method}{7}
\contentsline {section}{\numberline {3}Forward flux sampling}{8}
\contentsline {subsection}{\numberline {3.1}FFS variants}{10}
\contentsline {subsubsection}{\numberline {3.1.1}{{``Direct'' FFS}}}{10}
\contentsline {subsubsection}{\numberline {3.1.2}{{Branched Growth}}}{11}
\contentsline {subsubsection}{\numberline {3.1.3}{{The ``Rosenbluth-like'' method}}}{12}
\contentsline {subsection}{\numberline {3.2}Requirement for stochastic dynamics}{13}
\contentsline {subsection}{\numberline {3.3}Pruning}{14}
\contentsline {section}{\numberline {4}Computational efficiency: prediction and optimisation}{14}
\contentsline {subsection}{\numberline {4.1}Analytical expressions for the efficiency}{14}
\contentsline {subsection}{\numberline {4.2}Optimising the efficiency}{17}
\contentsline {subsubsection}{\numberline {4.2.1}Optimising the number of trial runs}{17}
\contentsline {subsubsection}{\numberline {4.2.2}Optimising the interface positions}{18}
\contentsline {subsubsection}{\numberline {4.2.3}Efficiency gains}{19}
\contentsline {section}{\numberline {5}The order parameter and the committor}{19}
\contentsline {subsection}{\numberline {5.1}Defining the order parameter}{20}
\contentsline {subsection}{\numberline {5.2}The committor and the reaction coordinate}{21}
\contentsline {subsection}{\numberline {5.3}Using the committor to optimise the order parameter in FFS}{22}
\contentsline {section}{\numberline {6}Computing stationary distributions}{23}
\contentsline {subsection}{\numberline {6.1}Obtaining stationary distributions from FFS simulations}{24}
\contentsline {subsection}{\numberline {6.2}Forward flux / umbrella sampling}{26}
\contentsline {subsection}{\numberline {6.3}Nonequilibrium umbrella sampling in multiple dimensions}{26}
\contentsline {section}{\numberline {7}Applications}{27}
\contentsline {subsection}{\numberline {7.1}Genetic switch flipping}{27}
\contentsline {subsection}{\numberline {7.2}Homogeneous crystal/bubble nucleation}{31}
\contentsline {subsection}{\numberline {7.3}Nucleation in a sheared Ising model}{32}
\contentsline {section}{\numberline {8}Challenges and future directions}{34}


\section{Introduction}
``Rare events'' are fluctuation-driven transitions that have a low probability of occurring, but often  have important consequences when they do occur. Examples are ubiquitous, ranging from large-scale events such as earthquakes, global climate changes, financial crashes and telecommunications network failures, to smaller-scale processes typical of soft condensed matter and biological physics, such as activated chemical reactions, nucleation phenomena, protein conformational changes, switching in biochemical networks and translocation through pores. Computer simulation has an important role to play in understanding rare events, especially as they are often  difficult  to study experimentally. However, rare events are notoriously difficult to simulate, simply because in the typical simulation time few, if any, events happen. To address this issue, specialised techniques for simulating rare events have been developed in various different contexts over many years. Excellent reviews have already been published on this subject in the fields of condensed matter, chemical and biological physics \cite{daan,bolhuis_arpc,dellago2,dellago2007,vanerp2,dellago2008}. In this short topical review article, we focus only on one recently developed class of techniques, known as Forward Flux Sampling (FFS). Although this class of methods is still rather young (at least in this field), it has been applied to a variety of different problems, and several variants and improvements to the methodology have recently been proposed. A number of potential problems have also been highlighted. This article aims to bring together these developments, together with practical advice on using the methods and suggestions for directions of future research. In the late stages of preparation of this article, we became aware of an almost simultaneous review, also focusing on FFS, by Escobedo {\em{et al}} \cite{escobedo}. Although some duplication of material between these articles is inevitable, we hope to present a complementary perspective.

Typically, when studying a rare event process, one wishes to know how often the event happens, or equivalently the rate constant $k_{AB}$ for transitions from an initial state A to a final state  B. If the transition occurs between two time-invariant steady states and is itself is fast compared to $k_{AB}^{-1}$, the rate constant $k_{AB}$ will be time-invariant. The distribution function $F(T)$ for the time taken for an equilibrated configuration in the $A$ state to ``escape'' to the $B$ state is then given by:
\begin{equation}
F(T) = k_{AB}\, e^{-k_{AB}T}
\end{equation}
One is usually also interested in the mechanism by which the rare event process happens. For example, for a crystal nucleation problem, one might wish to know the crystal structure and shape of the growing nucleus, or for a protein folding problem, in what order the secondary structure elements form. Information on the mechanism can be obtained by sampling the  transition path ensemble (TPE), which is the ensemble of trajectories corresponding to transitions from $A$ to $B$. However, extracting simple and intuitive conclusions from these transition paths can sometimes be difficult. 

The FFS methodology discussed here was originally developed for simulations of rare events in nonequilibrium systems, although it can also be used for equilibrium systems. In this review, we consider ``equilibrium'' systems to be those whose dynamical rules obey detailed balance [regardless of whether they are actually in a stationary state]. Detailed balance has the consequence that for these systems, the stationary phase space probability distribution is given by the Boltzmann distribution \cite{daan}, and the system dynamics is time reversible. In contrast, nonequilibrium dynamical systems do not obey detailed balance, their stationary phase space distribution is not known {\em{a priori}} and their dynamics are not time reversible. Nonequilibrium systems present a host of important and interesting rare event processes.  However, these systems pose particular challenges for rare event simulation methods, as we shall discuss.

\section{Background}\label{sec:back}
In this section, we present a brief overview of rare event simulation methods in the area of condensed matter, chemical and biological physics. Our aim is to provide the background information necessary for the subsequent discussion of the FFS technique, rather than to give a comprehensive review. Consequently, some important methods are omitted completely, or only discussed very briefly, for which we apologise.
\subsection{``Bennett-Chandler'' type methods}
``Bennett-Chandler'', or ``reactive flux'' methods are based on the transition state theory (TST) expression for the rate constant \cite{eyring,wigner,horiuti}. In TST, phase space is partitioned by a dividing surface between ``reactant'' and ``product'' regions.  The rate constant is given by:
\begin{equation}\label{eq:tst}
k_{AB}^{TST} = \frac{\langle |\dot q |\rangle}{2 }\frac{ e^{-W(q^*) / (k_BT)}}{\int_{-\infty}^{q^*} dq e^{-W(q^*) / (k_BT)}}
\end{equation} 
where $q$ is an order parameter  that separates the reactant and product regions and measures the progress of the system between these regions, $q^*$ defines the dividing surface and $W(q)$ is the reversible work needed to move the system from state A to a value $q$ of the order parameter. This is proportional to the free energy, projected onto the coordinate $q$. The exponential term describes the equilibrium (Boltzmann) probability of finding the system at the dividing surface relative to the A state, while the term $\langle |\dot q |\rangle/2$ is the average velocity of the system from reactants to products across the dividing surface. Eq.(\ref{eq:tst}) assumes that all crossings of the dividing surface contribute to the rate constant, while  in reality a single trajectory can cross the dividing surface many times. The ``reactive flux'' formalism therefore corrects the TST expression with a ``transmission coefficient'' $\kappa$, which is less than unity \cite{bencha2,bencha1,dellago2}:
\begin{equation}\label{eq:tst2}
k_{AB} = \kappa(t) k_{AB}^{TST}
\end{equation}
The transmission coefficient $\kappa (t)$ replaces the factor $\langle |\dot q |\rangle/2$ in Eq.(\ref{eq:tst}) by the average initial velocity of trajectories, initiated from an equilibrium distribution at $q^*$, which are in the product basin after time $t$. For  times intermediate between the molecular timescale and the timescale for transitions between the reactant and product basins, $\kappa(t)$ is independent of time; for this reason we do not include an explicit time dependence for $k_{AB}$ in Eq.(\ref{eq:tst2}). In Bennett-Chandler type methods, one chooses an order parameter $q$ and computes the free energy profile $W(q)$ using a method such as Umbrella Sampling \cite{umbrella1,umbrella2,umbrella3}.  The transmission coefficient $\kappa$ is then computed by initiating a large number of trajectories from an equilibrium distribution restricted to $q=q^*$ (usually taken to be the maximum of $W(q)$),  and counting the fraction of these that end up in the product state. Bennett-Chandler-type methods are conceptually simple, easy to implement, and have been widely used. However, because of the assumption of the Boltzmann distribution inherent in Eq.(\ref{eq:tst}), these methods are not suitable for nonequilibrium systems. These methods also tend to be rather sensitive to poor choices of the order parameter, since this will result in a small value of $\kappa$ which is hard to compute accurately. 

\subsection{Transition Path Sampling}
Transition Path Sampling (TPS) methods \cite{dellago1,dellago1999,dellago2,bolhuis_arpc} focus  directly on sampling the transition path ensemble (TPE) using a Monte-Carlo procedure in trajectory space. A single trajectory connecting the reactant and product regions of phase space is generated, and this is used to produce new trajectories. Several methods are available for generating new trajectories, of which probably the most widely used is ``shooting''. Here, a time-slice from the initial trajectory is then selected, and a small change is made, usually to the momentum coordinates. A new trajectory is then generated by integrating the system dynamics forward and backward in time from this slightly altered phase space point. If the new trajectory still joins the reactant and product basins, it is accepted into or rejected from a collection of computed transition paths with a probability that depends on its path weight, the weight of a path of length $n$ steps with phase space coordinates $\{ x\}$ being given by:
\begin{equation}\label{eq:tps}
{\mathcal{P}}[\{ x\}] = \rho(x_0)\prod_{i=0}^{n-1} p(x_{i}\to x_{i+1})
\end{equation}
where $\rho(x_0)$ is the phase space probability density for the initial point in the path and $p(x_{i}\to x_{i+1})$ is the probability of making a simulation step from $x_{i}$ to $x_{i+1}$. In practice, for Molecular Dynamics simulations in the NVE ensemble, one can simply accept all generated paths that connect the reactant and product basins. However, because Eq.(\ref{eq:tps}) requires knowledge of the phase space distribution $\rho_0$, TPS is not suitable for nonequilibrium systems (although a TPS method for nonequilibrium systems has been proposed  \cite{crooks2001}).

Computation of the rate constant $k_{AB}$ in TPS is based on the correlation function $C(t)$ \cite{dellago2}:
\begin{equation}\label{eq:ct}
C(t) = \frac{\langle h_A(x_0) h_B(x_t)\rangle}{\langle h_A(x_0)\rangle}
\end{equation}
where $h_A(x)$ is unity in the reactant basin and zero elsewhere, and $h_B(x)$ is unity in the product basin and zero elsewhere. $C(t)$ is the probability of finding the system in the product basin at time $t$, given that at time $0$ it was in the reactant basin. For times longer than the molecular timescale, $C(t) \approx k_{AB}t$. In practice, one computes $C(t)$ in two stages: a TPS simulation between reactant and product regions, and an  ``umbrella sampling'' procedure in which the end points of transition paths are constrained to lie in a series of windows between A and B, defined by an order parameter. More information about TPS, and about the many improvements to the method which have been made, are given in  Refs \cite{dellago1,dellago1999,dellago2,bolhuis_arpc,dellago2007,dellago2008}. TPS has the advantage that it samples paths without the need for an order parameter, although an order parameter is needed to compute the rate constant.

\subsection{Transition Interface Sampling}\label{sec:tis}
Transition Interface Sampling (TIS) \cite{vanerp1,vanerp2} is a variant of TPS in which the rate constant is calculated differently. In TIS, phase space is divided up by a series of non-intersecting interfaces, defined by an order parameter $\lambda$, such that the reactant region is defined by $\lambda < \lambda_A = \lambda_0$ and the product region by $\lambda > \lambda_B = \lambda_n$. The expression for $k_{AB}$ used in TIS is then  \cite{vanerp1,vanerp2,anderson1995}:
\begin{equation}\label{eq:tis1}
k_{AB} = \frac{\overline{\Phi}_{A,n}}{\overline{h}_{\mathcal{A}}} = \frac{\overline{\Phi}_{A,0}}{\overline{h}_{\mathcal{A}}} P(\lambda_n | \lambda_0)
\end{equation}
where $\overline{\Phi}_{A,n}$ is the steady-state flux of trajectories leaving the $A$ state and reaching interface $\lambda_n$ ({\em{i.e.}} the $B$ state), and $h_{\mathcal{A}}$ is a history-dependent function that is unity if a trajectory was more recently in $A$ than in $B$, and zero otherwise. The right-hand side of Eq.(\ref{eq:tis1}) expresses the fact that the flux of trajectories that leave $A$ and cross $\lambda_n$ is equal to the flux of those leaving $A$ and crossing $\lambda_0$, multiplied by the probability $P(\lambda_n | \lambda_0)$ that a trajectory that crosses $\lambda_0$, coming from $A$, will subsequently reach $\lambda_n$ before returning to $A$. The flux $\overline{\Phi}_{A,0}$ is easy to calculate, since trajectories coming from $A$ cross $\lambda_0$ frequently. However, the probability $P(\lambda_n | \lambda_0)$ is small and thus difficult to calculate. This difficulty is overcome by expressing $P(\lambda_n | \lambda_0)$ as \cite{vanerp1}:
\begin{equation}\label{eq:tis2}
P(\lambda_n | \lambda_0) = \prod_{i=0}^{n-1}P(\lambda_{i+1} | \lambda_i)
\end{equation}
where the product is over all interfaces and $P(\lambda_{i+1} | \lambda_i)$ is the conditional probability that a trajectory that comes from $A$ and crosses $\lambda_i$ for the first time will subsequently reach $\lambda_{i+1}$ instead of returning to $A$. We note that expression (\ref{eq:tis2}) does not involve a Markovian approximation, because the probabilities $P(\lambda_{i+1} | \lambda_i)$ are conditional on the history of the trajectories reaching $\lambda_i$. Expressions (\ref{eq:tis1}) and (\ref{eq:tis2}) are known as the ``effective positive flux'' formulation of the rate constant.

In TIS, the flux $\overline{\Phi}_{A,0}$ is computed using a ``brute-force'' simulation in the $A$ basin. TPS is then used to sample  the ensemble of transition paths from the reactant basin to $\lambda_i$  (using as an initial path one of the successful paths to $\lambda_{i-1}$).  The fraction of transition paths ultimately reaching $\lambda_{i+1}$, as opposed to $\lambda_A$, in the ensemble of paths from $A$ to $\lambda_i$, is an estimate for $P(\lambda_{i+1} | \lambda_i)$. It is important to recognise that in TIS, the interfaces are simply used as a convenient way of dividing the transition paths into sections. The order parameter need not correspond to the true reaction coordinate and it is not assumed that the system loses its ``memory'', or becomes uncorrelated, between one interface and the next. However, a version of TIS in which one does assume decorrelation between interfaces,  Partial Path Transition Interface Sampling (PPTIS) \cite{moroni1}, is more efficient for diffusive transitions. A number of improvements to TIS have been developed in recent years, including computation of free-energy barriers \cite{Moroni3}, swapping partial paths between ensembles at different interfaces \cite{vanerp2,vanerp2007,vanerp2008} and sampling transitions to multiple final states \cite{jutta}.

\subsection{Milestoning}

The ``Milestoning'' method \cite{faradjian04,west07} also uses a series of interfaces between the initial and final states, defined by an order parameter $\lambda$. In contrast to TIS, milestoning does assume memory loss between interfaces. Short simulation trajectories are initiated from quasi-equilibrium (or first hitting point \cite{ve08}) distributions at interface $\lambda_i$, and continued until they reach the adjacent interfaces $\lambda_{i-1}$ or  $\lambda_{i+1}$. From these trajectories, first passage time distributions for transitions between interfaces are obtained, and these can be used to compute the time evolution of the system. In contrast to the other methods discussed above (and FFS), Milestoning does not assume that the transition is a simple two-state process with exponential kinetics. A new variant of this method, ``Markovian milestoning'', has recently been proposed \cite{ve3}: here, the interfaces are defined by Voronoi polyhedra (see Section \ref{sec:voronoi}) and the kinetic information is obtained by running trajectories confined to these interfaces by reflecting boundary conditions.

\subsection{The Weighted Ensemble method}
The Weighted Ensemble method of Huber and Kim \cite{huber1996} is also an interface-based method, and is closely related to FFS. This method  divides the phase space region between the reactant and product states into a series of bins, and simulates a collection of ``walkers'', each of which carries a probability weight, and which either merge or divide as they progress between bins, so as to maintain the number of walkers in each bin. By monitoring the flux of walkers across the interfaces the transition rate constant can be efficiently computed. 

\subsection{The Finite Temperature String Method}
The Finite Temperature String method (FTS) is different in concept to the above methods, since it focuses on the ``principal curve'' between A and B \cite{e_05}. This is the path that follows the averaged position of the system, projected onto a series of hyperplanes perpendicular to the path itself. For systems with overdamped Langevin dynamics, the free energy along the principal curve can be directly related to the committor function \cite{e_05}.  The FTS method defines a string of ``beads'', or configurations, between A and B, and iteratively refines the positions of the beads until the string corresponds to the principal curve. In the original version of FTS, this was achieved by performing constrained simulations on the hyperplanes perpendicular to the string. However, a simplified version has recently been published \cite{Venturoli2} in which one instead defines Voronoi polyhedra around the beads (see Section \ref{sec:voronoi}) and simulates multiple copies of the system, each one constrained to lie inside one of the Voronoi polyhedra. The beads then evolve towards the average configuration obtained in the simulation for that polyhedron.

\section{Forward flux sampling}\label{sec:FFS}
Forward flux sampling (FFS) methods \cite{FFS,FFS2} were developed to simulate rare events in nonequilibrium systems with stochastic dynamics. Nonequilibrium systems are ubiquitous in condensed matter, chemical  and biological physics. However, their lack of detailed balance and the consequent absence of  a Boltzmann-like stationary distribution function and lack of time reversal symmetry mean that Bennett-Chandler-type methods, TPS, TIS, Milestoning, and most versions of the string method, are not suitable for these systems. FFS is thus one of only a few methods available for simulating rare events in nonequilibrium systems. For equilibrium systems, FFS provides an alternative to the above methods. The development of FFS was inspired by TIS. However, we have subsequently become aware of the prior existence of a similar class of methods used in telecommunications modelling, known as RESTART \cite{kahn,altamirano,bayes,altamirano2,altamirano3}.

FFS, like TIS,  uses a series of interfaces between the initial and final states to calculate the transition rate and to sample the transition path ensemble. These interfaces are defined by an order parameter  $\lambda$:  the initial (A) state is defined by $\lambda < \lambda_A = \lambda_0$, the final (B) state by $\lambda > \lambda_B = \lambda_n$, and the remaining interfaces are defined by intermediate values of $\lambda$: $\{\lambda_i \dots \lambda_{n-1}\}$. The method requires that  $\lambda_{i+1} > \lambda_i$ for all $i$, and that any trajectory from A to B passes through each interface in turn. This places no restriction on the trajectories, which are free to loop back to recross previous interfaces any number of times. Like TIS, FFS  uses the effective positive flux expression for the rate constant, Eqs. (\ref{eq:tis1}) and (\ref{eq:tis2}). However, FFS differs fundamentally from TIS in the manner in which the conditional probabilities $P(\lambda_{i+1}|\lambda_i)$ and the transition paths are computed. In FFS,  the system dynamics are integrated forward in time only, eliminating the requirement for detailed balance.

Broadly speaking, FFS works by ``capitalising'' on fluctuations of the system dynamics in the direction of the order parameter. When the system undergoes a fluctuation that reaches the first interface, its configuration is stored. This stored configuration is then used as the starting point for repeated ``trial runs'', to evaluate the probability that the system will reach the next interface. These trial runs are continued until the system reaches either the next interface (a ``success''), or returns to the A state (a ``failure''). The end points of successful trials are used to initiate new trial runs, to the subsequent interface. The result is that the system is driven in a ratchet-like manner from the initial to the final state, without imposing any bias on the microscopic  dynamics. The probabilities $P(\lambda_{i+1}|\lambda_i)$ of Eqs. (\ref{eq:tis1}) and (\ref{eq:tis2}) are obtained from the fraction of successful trial runs at each interface, and these can be multiplied by the flux of trajectories crossing the first interface to obtain the rate constant $k_{AB}$. A correctly weighted collection of transition paths is obtained by tracing back trial runs from the final state to the initial state \cite{FFS2}. Because each trial run starts from the final point of a previous trial run, the correct system dynamics is preserved along the whole transition path.  We note that although FFS does assume that transitions between A and B are uncorrelated and that the rate $k_{AB}$ is time-invariant, there is no requirement for the B state to be stable. Although FFS has generally been used for systems with stable A and B states, it can also be used to predict the probability of rare fluctuations from a stable A state, along a chosen order parameter, to an arbitrarily positioned end point $\lambda_n$ \cite{borrero2009}. It is also important to note that FFS is a static sampling technique, in which each new transition path is generated from scratch. This is in contrast to dynamic methods such as TPS and TIS, in which new transition paths are generated by modifying already existing ones. The advantage of static methods is that they generate uncorrelated samples, making them likely to explore a wider range of path space. However, static methods may also waste computational effort by repeatedly sampling blind alleys.

\begin{figure}[h!]
\begin{center}
{\rotatebox{0}{{\includegraphics[scale=0.4,clip=true]{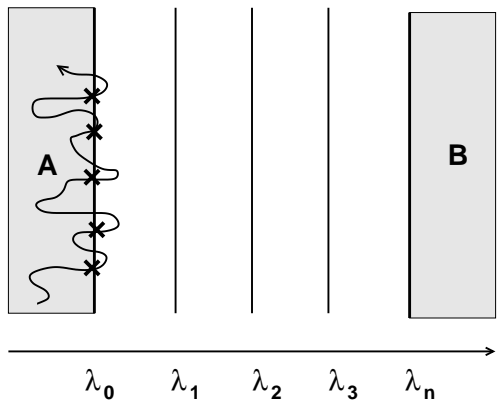}}}}
\caption{Schematic illustration of the initial simulation in the A state. Configurations corresponding to the points in the trajectory marked with crosses are stored.\label{fig:ffs2} }
\end{center}
\end{figure}

Within the FFS framework, various different protocols have been proposed for firing trial runs and storing configurations at the interfaces. All these variants begin with a simulation in the A state, illustrated schematically in Figure \ref{fig:ffs2}. This simulation is used to compute the flux $\overline{\Phi}_{A,0}/\overline{h}_{\mathcal{A}}$ of trajectories out of the A state [Eq.(\ref{eq:tis1})], as well as to obtain a sample of configurations corresponding to crossings of interface $\lambda_0$, to act as starting points for the trial run procedure. The system is initiated in the A state, and (after an initial equilibration period) the time evolution of the order parameter $\lambda$ is monitored. When the system crosses $\lambda_0$ coming from A ({\em{i.e.}} in the direction of increasing $\lambda$), a counter is incremented, and the system configuration is stored. The simulation is then continued until $N_0$ configurations have been stored. The flux  $\overline{\Phi}_{A,0}/\overline{h}_{\mathcal{A}}$ is obtained by dividing the number of crossings $N_0$ by the total simulation time (which includes time spent on fluctuations away from the A state but does not include time spent in the B state \cite{FFS2}).

The FFS variants differ in the way the probability $P(\lambda_n | \lambda_0)$  is computed. In our original presentation of FFS, Refs \cite{FFS} and \cite{FFS2}, we described three different variants. These are outlined below. Since Ref \cite{FFS2}, more advanced versions of some of these algorithms have been proposed \cite{borrero2007,borrero2008,borrero2009}. These advances are discussed in Sections \ref{sec:eff}, \ref{sec:op} and \ref{sec:sd}.

\subsection{FFS variants}\label{sec:variants}
\begin{figure}[h!]
\begin{center}
{\rotatebox{0}{{\includegraphics[scale=0.4,clip=true]{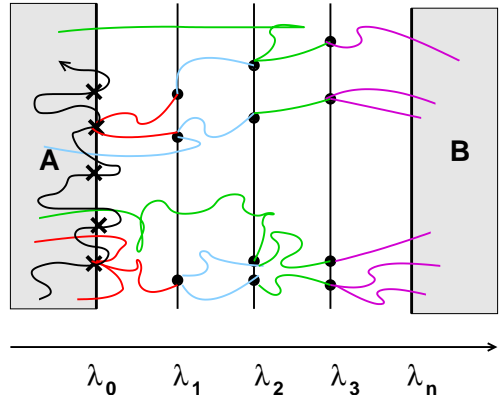}}}}
\caption{Schematic illustration of the DFFS method. An ensemble of branched transition paths is generated simultaneously by firing $M_i$ trial runs from randomly chosen configurations at each interface in turn. The different colours denote the trial runs fired in the different stages of the procedure (in the order red, blue, green, purple). \label{fig:dffs} }
\end{center}
\end{figure}

\subsubsection{{{``Direct'' FFS}}} The original version of FFS \cite{FFS,FFS2} has subsequently been termed ``Direct-FFS'', or DFFS \cite{borrero2007}. In this algorithm, many transition paths are generated simultaneously, using the following procedure (illustrated in Figure \ref{fig:dffs}):
\begin{enumerate}
\item{Carry out a simulation in the A basin to generate a collection of $N_0$ configurations corresponding to crossings of interface $\lambda_0$, as well as an estimate of the flux  $\overline{\Phi}_{A,0}/\overline{h}_{\mathcal{A}}$.} 
\item{Choose a configuration from this collection at random and use it to initiate a trial run which is continued until it either reaches $\lambda_1$ or returns to $\lambda_0$. If $\lambda_1$ is reached, store the end point of the trial run. Repeat this $M_0$ times, each time choosing a random starting configuration from the collection at $\lambda_0$. Compute $P(\lambda_1|\lambda_0)$ from the fraction of successful trial runs.}
\item{Repeat step (ii) using the stored configurations at $\lambda_1$ to initiate $M_1$ trial runs to $\lambda_2$ or back to $\lambda_0$. Generate a new collection of configurations  at $\lambda_2$ from the end points of successful trials. Estimate  $P(\lambda_2|\lambda_1)$ from the fraction of successful trials.}
\item{Repeat until $\lambda_n$ is reached.}
\end{enumerate}
Statistical errors can be computed by repeating the DFFS procedure several times or using analytical expressions (see Section \ref{sec:eff}). DFFS is straightforward to implement and its computational efficiency is rather robust to the choice of parameters (see Section \ref{sec:eff}), although it does require the storage of many configurations at each interface, which may be an issue for large-scale simulations. In order to extract transition paths from a DFFS simulation, one needs to record the connectivity history of all the trial runs. This allows one to piece together {\em{a posteriori}} complete transition paths from the full set of stored trial runs.  Finally, it is important to note that the transition paths generated by DFFS are branched: many paths may start from a single configuration at $\lambda_0$. 

\subsubsection{{{Branched Growth}}}
\begin{figure}[h!]
\begin{center}
{\rotatebox{0}{{\includegraphics[scale=0.4,clip=true]{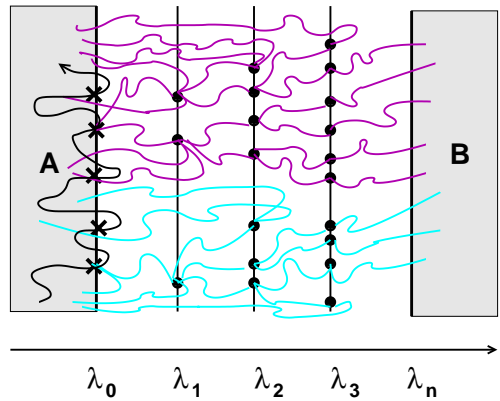}}}}
\caption{Schematic illustration of the BG algorithm. Branched transition paths are generated by firing $k_{i}$ trials from the end points at $\lambda_i$ of all successful trials from $\lambda_{i-1}$. The colours denote different branched paths, which are generated sequentially.  \label{fig:bg} }
\end{center}
\end{figure}

The ``Branched Growth'' (BG) algorithm \cite{FFS2} generates branched transition paths from A to B one at a time, rather than simultaneously as in DFFS. The algorithm proceeds as follows (see Figure \ref{fig:bg}):
\begin{enumerate}
\item{Evaluate  $\overline{\Phi}_{A,0}/\overline{h}_{\mathcal{A}}$ and generate configurations at $\lambda_0$ using a simulation in the A basin.}
\item{For the first configuration at $\lambda_0$, fire $k_0$ trial runs, which are continued until $\lambda_1$ or $\lambda_0$ is reached. Store the end configurations of all successful trial runs.}
\item{From each of these stored points at $\lambda_1$, initiate $k_1$ trial runs to $\lambda_2$ or back to $\lambda_0$. Store the end points of successful trials.}
\item{Iterate this procedure over all subsequent interfaces until B is reached, or until no trials are successful at a given interface. This generates a branching tree of paths all starting from the same configuration at $\lambda_0$. Estimate $P(\lambda_{i+1} | \lambda_i)$ as the total number of trials to reach B, divided by the total possible number $\prod_{i=0}^{n-1}k_i$.}
\item{Repeat steps (ii) to (iv) for subsequent configurations at $\lambda_0$ and average the estimate for $P_B$ over many path generations (note  that any zero values should be included in the average).}
\end{enumerate}
The BG method can easily be coded as a recursive algorithm, and has the potential advantage that it does not require storage of large numbers of configurations at each interface, and that extraction of transition paths is simpler than for DFFS (since these are generated one at a time). The BG method is however rather sensitive to parameter choice (see Section \ref{sec:eff}). If the number of trial runs per interface is too large, the method generates highly branched transitions paths, so that sampling the later interfaces is computationally expensive. If too few trials are chosen per interface, few paths succeed in reaching the later interfaces. However, Borrero and Escobedo have proposed a method for automatic optimisation of the parameters \cite{borrero2008}, which is discussed in Section \ref{sec:eff}. The same authors have used the BG method as the basis for  the FFS-LSE method \cite{borrero2007}, discussed in Section \ref{sec:lse}.

\subsubsection{{{The ``Rosenbluth-like'' method}}} 

\begin{figure}[h!]
\begin{center}
{\rotatebox{0}{{\includegraphics[scale=0.4,clip=true]{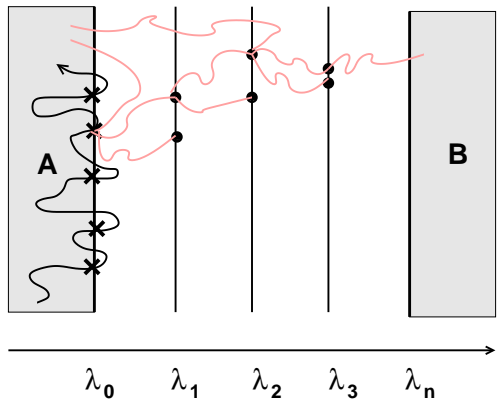}}}}
\caption{Schematic illustration of the RB method. Unbranched transition paths are generated one at a time by firing $k_{i}$ trials from one randomly chosen  end point at $\lambda_i$ from the successful trials fired from $\lambda_{i-1}$.  \label{fig:rb} }
\end{center}
\end{figure}

The Rosenbluth-like (RB) FFS variant allows the sequential generation of unbranched transition paths \cite{FFS2}. It draws on a close analogy between FFS and the sampling of polymer configurations in Monte Carlo simulations. In a polymer simulation, one seeks to grow a new polymer chain, monomer by monomer, in an environment crowded with other polymers. This is analogous to the interface-by-interface growth of a transition path in FFS.  The RB FFS variant is an application of the Rosenbluth polymer sampling method \cite{rosenbluth,daan} to rare event simulations. The procedure is as follows:
\begin{enumerate}
\item{Evaluate  $\overline{\Phi}_{A,0}/\overline{h}_{\mathcal{A}}$ and generate configurations at $\lambda_0$ using a simulation in the A basin.}
\item{For the first configuration at $\lambda_0$, fire $k_0$ trial runs to $\lambda_1$ or $\lambda_0$. Store the end points of successful trials.}
\item{Choose at random one of the stored configurations at $\lambda_1$. Use this as the starting point for $k_1$ trial runs to $\lambda_2$ (or back to $\lambda_0$).}
\item{Iterate this procedure over all interfaces until $B$ is reached, or until no trials are successful.}
\item{Repeat steps (i) to (iv) for successive configurations at $\lambda_0$.}
\item{Compute $P(\lambda_{i+1} | \lambda_i)$ for each interface using a weighted average as described below. }
\end{enumerate}
In the RB method, the paths that are generated do not all have equal statistical weight. The weight $ w_{i,b}$  of  path $b$  from A to $\lambda_i$ is given by
\begin{equation}
w_{i,b} = \prod_{j=0}^{i-1}S_{j,b}/k_j
\end{equation}
where $S_{j,b}$ is the number of successful trials fired at interface $j$ during the generation of path $b$. To compute the probabilities $P(\lambda_{i+1}|\lambda_i)$, a weighted average is needed for each interface:
\begin{equation}\label{eq:rb1}
P(\lambda_{i+1}|\lambda_i) = \frac{\sum_{b} w_{i,b}S_{i,b}/k_i}{\sum_b w_{i,b}}
\end{equation}
Here, the index $b$ labels a specific path leading from A to a configuration at interface $\lambda_i$, and $S_{i,b}$ is the number of successful trials fired from that configuration to $\lambda_{i+1}$. When sampling over many transition paths, both the numerator and denominator of Eq.(\ref{eq:rb1}) will become large. In this case, Eq.(\ref{eq:rb1}) may become unsatisfactory, and one may prefer to reweight the paths  using a Metropolis acceptance/rejection procedure \cite{daan}, or alternatively a ``waste-recycling'' scheme \cite{wr} at each interface. This is described in detail in Ref \cite{FFS2}: however, we prefer simply to use Eq.(\ref{eq:rb1}),  since the approaches described in Ref.\cite{FFS2} are rather complicated to code. When analysing the properties of the transition paths  generated in the RB method, it is also necessary to include the weighting factor $w_n$. 

  The major advantage of the RB method is that the resulting transition paths are unbranched and sequentially generated, making them easy to extract and analyse. The RB method is also easy to code as a recursive algorithm. 

\subsection{Requirement for stochastic dynamics}
FFS requires the dynamics of the system to be stochastic, since for deterministic dynamics all trial runs fired from a given configuration at interface $\lambda_i$ would be identical. However, this requirement allows for a wide range of possibilities, including kinetic Monte Carlo simulations of chemical reaction networks \cite{FFS,morelli_jcp,morelli_bpj,morelli_lambda}, lattice models with Monte Carlo spin flips \cite{ising_shear1,ising_shear2,page2006} and particle-based Brownian dynamics or Monte Carlo simulations \cite{FFS2,sanz2007,vanzon}. FFS is not suitable for use with completely deterministic Molecular Dynamics (MD) algorithms; however, it has been successfully applied to MD simulations by including a weak coupling to a stochastic Lowe-Andersen thermostat, which preserves the momentum of the system \cite{wang,jurasczek,vega}. Given that MD trajectories with infinitesimally different initial conditions diverge within a few picoseconds due to the Lyapunov instability, this is unlikely to constitute a severe perturbation to the system dynamics.

\subsection{Pruning}\label{sec:prune}
FFS as described above requires that trial runs fired from interface $\lambda_i$ be integrated until they reach $\lambda_{i+1}$ or return {\em{all the way back to A}}.  If transition paths are short, the computational cost of integrating failed trials back to $\lambda_0$ is likely to be rather low. However, in some cases, such as diffusive barrier crossings or intermediate metastable states, it may be expensive to integrate all the way back to $\lambda_0$. In these cases one can use a pruning scheme in combination with any of the above FFS variants. Here, trial runs from $\lambda_i$ are integrated only as far back as some pre-defined value $\lambda_p < \lambda_i$; typically, $\lambda_p = \lambda_{i-1}$. With some probability $p_{p}$, a trial run which reaches $\lambda_p$ is terminated and considered to have failed. If, with probability $(1-p_p)$, the trial run is not terminated, then its statistical weight is increased by a factor $f_p=1/(1-p_p)$. This requires minor modifications to be made to the three FFS algorithms described above, to include differential weights for configurations at $\lambda_i$. These are described in detail in Ref. \cite{FFS2}; however, we did not find that pruning produced much improvement in computational efficiency for the examples tested (a model genetic switch and a simple representation of polymer translocation) \cite{FFS2}. This approach may nevertheless prove useful for other problems.

\section{Computational efficiency: prediction and optimisation}\label{sec:eff}
An essential requirement for a rare event simulation method is that it should provide the rate constant and transition path ensemble more efficiently than brute force simulation. Defining, quantifying and optimising the computational efficiency of such methods is therefore an important task.  FFS involves a number of parameters: the number and position of the interfaces, the number of trials fired from each interface, the number of configurations stored at the first interface (for DFFS), as well as the choice of which FFS variant to use and the definition of the order parameter. The computed rate constant and transition paths should not (in principle) depend on any of these choices, but they will affect the efficiency. We use the simple definition for the computational efficiency $\mathcal{E}$ \cite{mooij}:
\begin{equation}\label{eq:eff}
\mathcal{E} = \frac{1}{\mathcal{CV}}
\end{equation}
where ${\mathcal{C}}$ is the computational cost (in simulation steps) of calculating the rate constant, and ${\mathcal{V}}$ is the statistical variance in the result, normalised by the square of its mean. A slightly different, but equivalent, expression was used by Van Erp in his analysis of the efficiency of TPS/TIS in comparison to Bennett-Chandler-type methods \cite{vanerp2006}.

\subsection{Analytical expressions for the efficiency}\label{sec:anal}
Analytical expressions for the efficiency of a method \cite{vanerp2006,FFS3} are  useful for several reasons. Firstly, they allow one to estimate, before beginning a lengthy calculation, how much effort will be required to obtain a desired level of accuracy. Secondly, they allow the estimation of error bars on a computed result, where it is not feasible to repeat the calculation. Thirdly, one can use the analytical expressions to optimise the efficiency of the method with respect to  parameter choice. It is possible to derive approximate expressions for the efficiency of the FFS methods discussed above, as  a function of the number of interfaces, the probability of success at each interface,  and the number of trials at each interface. Here, we simply present the key results; more details including derivations can be found in Ref. \cite{FFS3}.

For the subsequent discussion, it is important to note that in the BG and RB variants, the parameter defining the number of trials is $k_i$, the number of trials per configuration at $\lambda_i$. For DFFS, the relevant parameter is $M_i$, the {\em{total}} number of trials fired at interface $i$. To simplify our discussion, however, we shall use the notation $k_i$ also for DFFS, but in this context we define it as $k_i \equiv M_i/N_0$. We shall also simplify our notation for the probabilities, defining  $p_i \equiv P(\lambda_{i+1}|\lambda_i)$ and $q_i \equiv 1-p_i$.

The computational cost  ${\mathcal{C}}$ can be estimated by assuming that the average length of a trial run from $\lambda_i$ to  $\lambda_j$ is linearly proportional to   $|\lambda_i-\lambda_j|$, so that the average cost $C_i$ of a trial run fired from  $\lambda_i$ is \cite{FFS3}:
\begin{equation}\label{eq:ci}
C_i = Q[p_i (\lambda_{i+1}-\lambda_i) + q_i(\lambda_{i}-\lambda_A)]
\end{equation}
where  Q is a constant.  Eq.(\ref{eq:ci}) can be used to write down expressions for ${\mathcal{C}}$ for the three FFS variants discussed in Section \ref{sec:FFS}; these expressions differ because the variants differ in the average number of trials fired per starting point at $\lambda_0$. The results are:
\begin{equation}\label{eq:cdffs}
{\mathcal{C}}^{DFFS}= R + k_0C_0 + \sum_{i=1}^{n-1}\left[ k_i C_i \prod_{j=0}^{i-1}(1-q_j^{N_0k_j})\right]  \approx  R + \sum_{i=1}^{n-1}k_iC_i
\end{equation}
\begin{equation}\label{eq:cbg}
{\mathcal{C}}^{BG}= R + k_0C_0 + \sum_{i=1}^{n-1}\left[ k_i C_i \prod_{j=0}^{i-1}p_jk_j\right]
\end{equation}
and
\begin{equation}\label{eq:crb}
{\mathcal{C}}^{RB}= R + k_0C_0 + \sum_{i=1}^{n-1}\left[ k_i C_i \prod_{j=0}^{i-1}(1-q_j^{k_j}))\right]
\end{equation}
where the cost is defined per starting configuration at $\lambda_{0}$ and $R$ is the cost of generating such a starting configuration. These expressions take into account the fact that if no trials are successful at a given interface, the FFS algorithm will not make it to later interfaces.

The relative variance ${\mathcal{V}}$ in the computed rate constant is assumed to arise only from the computation of $P_B$ and not from the initial flux calculation. This is justified as long as the initial flux is large enough, and the initial simulation run in the $A$ basin is long enough.  The key assumption made in calculating ${\mathcal{V}}$ is that trial runs at subsequent interfaces are uncorrelated. This allows us to treat the number of successful trial runs from interface $i$ as a binomially distributed random variable, with parameter $p_i$. Taking into account the details of the different sampling protocols, we arrive at:
\begin{equation}\label{eq:vdffs}
{\mathcal{V}}^{DFFS}= \sum_{i=1}^{n-1}\frac{q_i}{p_ik_i}\left[ \frac{1}{\prod_{j=0}^{i-1}(1-q_j^{N_0k_j})}\right]\approx \sum_{i=1}^{n-1}\frac{q_i}{p_ik_i}
\end{equation}
and
\begin{equation}\label{eq:vbg}
{\mathcal{V}}^{BG}= \sum_{i=1}^{n-1}\frac{q_i}{\prod_{j=0}^{i}p_jk_j}.
\end{equation}
An equivalent expression was also derived in Ref. \cite{FFS3} for the RB method with Metropolis reweighting, but it is rather complicated. A much simpler expression can be derived if the $p_i$ values are instead computed using Eq.(\ref{eq:rb1}); this will be discussed in a forthcoming publication \cite{anderson}.

Expressions (\ref{eq:vdffs}) and (\ref{eq:vbg}) assume that the probability of success $p_i$ is the same for all trial runs fired from interface $i$. In reality, however, some configurations at $\lambda_i$ will have higher probability of success than others. This can be included in the expressions for ${\mathcal{V}}$ by assuming that the $p_i$s have an intrinsic, ``landscape variance'' $U_i$. This leads to minor modifications to the results: for details see Ref. \cite{FFS3}. Interestingly, the three FFS variants cope differently with this landscape variance. Because the DFFS and BG methods produce branched paths, they sample many configurations at interfaces $i>0$ as the number of trials $k_i$ becomes large. This makes them insensitive to the values of the landscape variance $U_i$ for $i>0$. By contrast, in the RB method, where the paths are not branched, only one configuration is sampled per interface per path, so that all the landscape variance values $U_i$ contribute to the total variance ${\mathcal{V}}$.

\begin{figure}[h!]
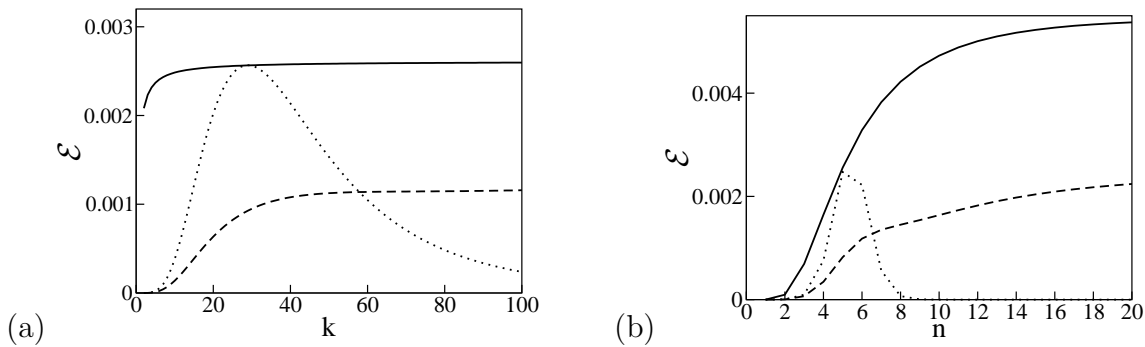

\begin{center}
\makebox[20pt][l]{(a)}{\rotatebox{0}{{\includegraphics[scale=0.25,clip=true]{allen_etal_fig5a.eps}}}}\hspace{1cm}\makebox[20pt][l]{(b)}{\rotatebox{0}{{\includegraphics[scale=0.25,clip=true]{allen_etal_fig5b.eps}}}}
\caption{Predicted computational efficiency ${\mathcal{E}}$ for the three FFS variants DFFS (solid lines), BG (dotted lines) and RB (with Metropolis acceptance/rejection; dashed lines), for a hypothetical rare event problem with evenly spaced interfaces and $p_i=p$, with $P_B=10^{-8}$, $Q=R$, $U_i=0$ and $N_0=1000$. (a): ${\mathcal{E}}$ as a function of the number of trials $k$, with 5 interfaces. (b): ${\mathcal{E}}$ as a function of the number of interfaces $n$, for $k=25$. [Reproduced with permission from Ref \cite{FFS3}].\label{fig:eff1} }
\end{center}
\end{figure}

The efficiency ${\mathcal{E}}$ is obtained by substituting Eqs. (\ref{eq:cdffs}-\ref{eq:vbg}) into Eq.(\ref{eq:eff}). Figure \ref{fig:eff1} shows ${\mathcal{E}}$ plotted as a function of the number of trials $k$ [panel (a)] and the number of interfaces $n$ [panel (b)] for a hypothetical rare event problem with evenly spaced interfaces and $p_i=p$ for all interfaces, with $P_B=10^{-8}$, $Q=R$ and $N_0=1000$. The striking result is that while the BG method is rather sensitive to the choice of either $n$ or $k$, both the DFFS and RB methods are insensitive to both these parameters, as long as $k$ and $n$ are large enough. This implies that the computational cost associated with having many interfaces or firing many trial runs is balanced by a proportional gain in statistical accuracy. Of course, this analysis cannot be taken to extremes: a very large number of interfaces will have associated overhead costs, as well as leading to strong correlations between successive interfaces, neither of which is taken into account in this analysis.

\begin{figure}[h!]
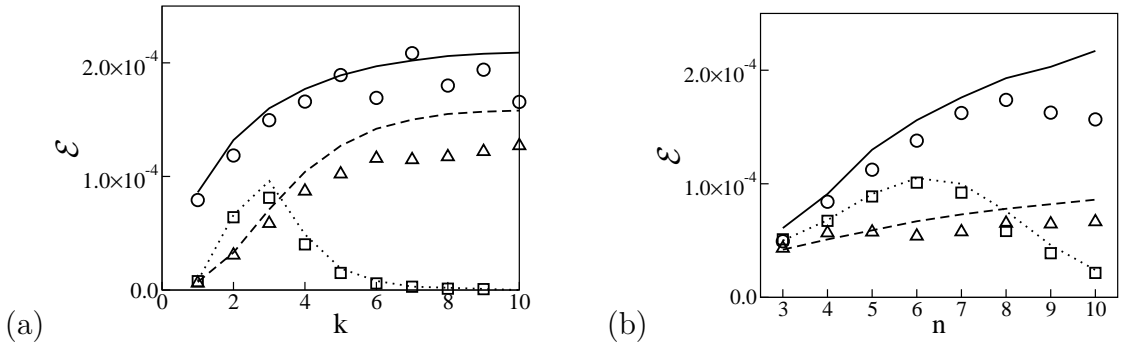

\begin{center}
\makebox[20pt][l]{(a)}{\rotatebox{0}{{\includegraphics[scale=0.25,clip=true]{allen_etal_fig6a.eps}}}}\hspace{1cm}\makebox[20pt][l]{(b)}{\rotatebox{0}{{\includegraphics[scale=0.25,clip=true]{allen_etal_fig6b.eps}}}}
\caption{Predicted  and measured efficiency ${\mathcal{E}}$ for the Maier-Stein system, simulated with overdamped Langevin dynamics [for details see Ref \cite{FFS3}]. Lines: predicted efficiency. Solid: DFFS, dotted: RB (with Metropolis acceptance/rejection), dashed: BG. Symbols: measured efficiency. Circles: DFFS, triangles: RB (Metropolis), squares: BG. (a):  ${\mathcal{E}}$ as a function of number of trials $k$ (b):  ${\mathcal{E}}$ as a function of number of interfaces $n$. [Reproduced with permission from Ref \cite{FFS3}].\label{fig:eff2} }
\end{center}
\end{figure}

Figure \ref{fig:eff2} shows the predicted efficiency ${\mathcal{E}}$, compared with the actual value measured for FFS simulations of the two-dimensional nonequilibrium rare event problem proposed by Maier and Stein \cite{ms1,ms2,ms3}. This consists of a particle moving with overdamped Langevin dynamics in a force field that is not the gradient of a potential field (for details see Ref. \cite{FFS3}). The values of $R$, $Q$, $\{p_i\}$ and $\{U_i\}$ were extracted from the FFS simulations and used as inputs to the analytical expressions. Not only do the results for the Maier-Stein system closely resemble the trends in Figure \ref{fig:eff1} for the hypothetical problem, but the agreement between the analytical results and the simulations is extremely good. However, such good agreement cannot be relied on in all cases: for a model genetic switch, where correlations between successive interfaces are more likely, differences of up to a factor of 10 between the analytical predictions for ${\mathcal{E}}$ and the simulation results were observed \cite{FFS3}.

\subsection{Optimising the efficiency}
Borrero and Escobedo \cite{borrero2008} have shown how these analytical expressions can be used to  optimise the parameters in FFS simulations, for the DFFS and BG schemes. They adopt two complementary approaches: (i) optimising the number of trial runs $ \{k_i\}$ for a fixed set of interfaces, and (ii) optimising the positioning of the interfaces $ \{\lambda_i\}$, for fixed $ \{k_i\}$.

\subsubsection{Optimising the number of trial runs}
For a given set of interfaces, the optimum values for the number of trials $ \{k_i\}$ [or for DFFS, $ \{M_i\} \equiv  \{N_0k_i\}$], can be found by minimising analytically the variance ${\mathcal{V}}$ with respect to the $ \{k_i\}$ (or $ \{M_i\}$). The cost ${\mathcal{C}}$ is constrained to a fixed value using a Lagrange multiplier. This leads to implicit expressions for the optimum $ \{k_i\}$ ((or $ \{M_i\}$) for the BG and DFFS schemes \cite{borrero2008}. For DFFS, under the assumption that the $M_i$ values are large, so that $q_i^{M_i} \approx 0$, this expression is
\begin{equation}\label{eq:msdffs}
M_i = \frac{P_B}{\sqrt(\alpha)}\left(\frac{1-p_i}{p_i}\right)^{1/2}\left(\frac{C_{i}}{Q} \right)^{-1/2}
\end{equation}
where $\alpha$ is the Lagrange multiplier that sets the cost, and $C_i$ and $Q$ are as defined in Section \ref{sec:anal}, so that $C_i/Q=[p_i (\lambda_{i+1}-\lambda_i) + q_i(\lambda_{i}-\lambda_A)]$. A simple practical procedure is then prescribed to obtain the optimum $\{M_i\}$ set, for a fixed cost, for DFFS:
\begin{enumerate}
\item{Set one of the  $M_i$ values (e.g. $M_0$).}
\item{Compute the other $M_i$s using:
\begin{equation}\label{eq:mmdffs}
\frac{M_{i+1}}{M_i}=  \left(\frac{p_{i}( \lambda_i- \lambda_A)}{p_{i+1}( \lambda_{i+1}- \lambda_A)}\right)^{1/2}
\end{equation}
(this assumes that $p_i$ is small; if not the expression is slightly more complicated).}
\item{Compute the cost associated with this $ \{M_i\}$ set from Eq.(\ref{eq:cdffs}).}
\item{Iterate to obtain a set of $M_i$s corresponding to the desired cost.}
\end{enumerate}
For the BG method, the equivalent expression to Eq.(\ref{eq:msdffs}) is more complicated but a similar principle applies; for details see Ref \cite{borrero2008}.

\subsubsection{Optimising the interface positions}
For a fixed set of $M_i$ (or $k_i$ values), the efficiency can be optimised with respect to the positions of the interfaces $ \{\lambda_i\}$, for $0 < i < n$. Borrero and Escobedo assume that the computational cost is fixed by the $ \{M_i\}$ (or $ \{k_i\}$) \cite{borrero2008}, and minimise the variance ${\mathcal{V}}$ with respect to the probabilities $ \{p_i\}$, with the constraint that $P_B=\prod_{i=0}^{n-1}p_i$ remain constant. This leads to the intuitive result that for optimum interface placement, there should be a constant flux of partial path trajectories across all interfaces. This implies that the product $M_ip_i$ (for DFFS), or $k_i \prod_{j=0}^{i-1}k_jp_j$ (for BG), should be constant across all interfaces. Since the $\{M_i\}$ (or $\{k_i\}$) are fixed, this specifies the optimum values for the probabilities $\{p_i\}$, which can be achieved by a suitable placement of the interfaces. To translate between $\{p_i\}$ values and the interface positions $\{\lambda_i\}$, one needs an interpolation function $f(\lambda_i)$, for which one choice is: 
\begin{equation}\label{eq:mmdffs_1}
f(\lambda_i) = \frac{\sum_{j=0}^{i-1}\ln{p_j}}{\sum_{j=0}^{n-1}\ln{p_j}}
\end{equation}
The optimisation procedure then consists of:
\begin{enumerate}
\item{Run FFS with an, as yet, non-optimal set of interfaces $\{\lambda_i\}$, to compute the function $f(\lambda_i)$.}
\item{Compute the optimum $\{p_i\}$ set by demanding constant flux across all interfaces.}
\item{Use the function $f(\lambda_i)$ to determine the interface placement corresponding to these optimal  $\{p_i\}$ values.}
\item{Iterate the procedure if necessary.}
\end{enumerate}
A separate methodology for optimising the position of the starting interface $\lambda_0$ has been proposed by Velez-Vega {\em{et al}} \cite{vega}.

\subsubsection{Efficiency gains}
\begin{figure}[h!]
\begin{center}
{\rotatebox{0}{{\includegraphics[scale=0.6,clip=true]{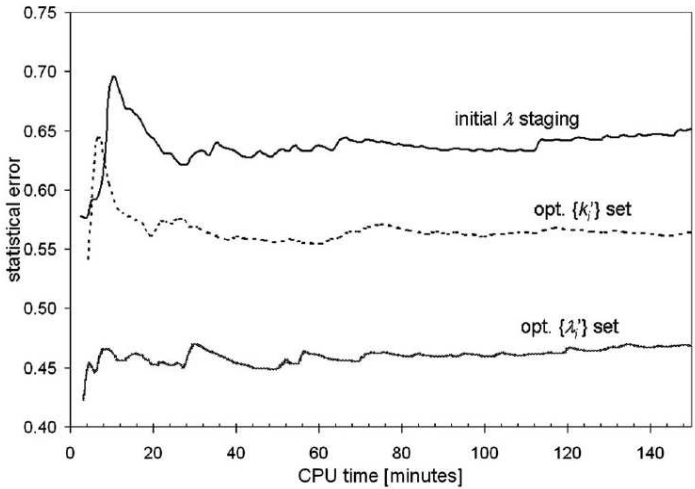}}}}
\caption{Statistical error in the estimated rate constant for BG FFS simulations of a two dimensional test potential with 4 interfaces, for an initially unoptimised set of interfaces (``initial $\lambda$ staging''), for the same set of interfaces with optimised number of trials (``opt. $\{k_i'\}$ set''), and for an optimised set of interface positions (``opt. $\{\lambda_i'\}$ set''). Smaller statistical errors are achieved for the same CPU time with the optimised parameters. [Reproduced with kind permission from Ref \cite{borrero2008}].\label{fig:borrero2} }
\end{center}
\end{figure}
Borrero and Escobedo demonstrate their efficiency optimisation procedure for a  simple two dimensional test potential, for the flipping of a model genetic switch (see Section \ref{sec:apps}) and for the folding of a lattice protein model \cite{borrero2008}, obtaining impressive results, as illustrated in Figure \ref{fig:borrero2}. For these examples, a single iteration proved to be enough to converge the optimisation for either the number of trials or the interface positions.

\section{The order parameter and the committor}\label{sec:op}

FFS relies on the definition of the order parameter $\lambda$, which must be some coordinate of the system that increases during the transition from A to B. FFS does not assume that $\lambda$ corresponds to the true ``reaction coordinate'', which is the actual route through phase space followed by the transition path ensemble. However, choosing a good order parameter (one which is close to the true reaction coordinate) will increase the computational efficiency, while a poor choice of order parameter will lead to wasted effort, as many of the configurations generated at interface $\lambda_i$ will have little chance of reaching $\lambda_{i+1}$. This may even lead to incorrect results if the number of paths sampled is small \cite{sear2008}. In this section, we first discuss how to define the order parameter in interface-based methods such as FFS. We then discuss the committor function, which contains information about the reaction mechanism, and which corresponds to the ``ideal'' choice of order parameter. Finally, we briefly review several methods which have recently been developed for extracting the reaction coordinate from measured committor values, focusing particularly on the FFS-LSE method of Borrero and Escobedo \cite{borrero2007}.

\subsection{Defining the order parameter}\label{sec:voronoi}
\begin{figure}[h!]
\begin{center}
\makebox[20pt][l]{(a)}{\rotatebox{0}{{\includegraphics[scale=0.4,clip=true]{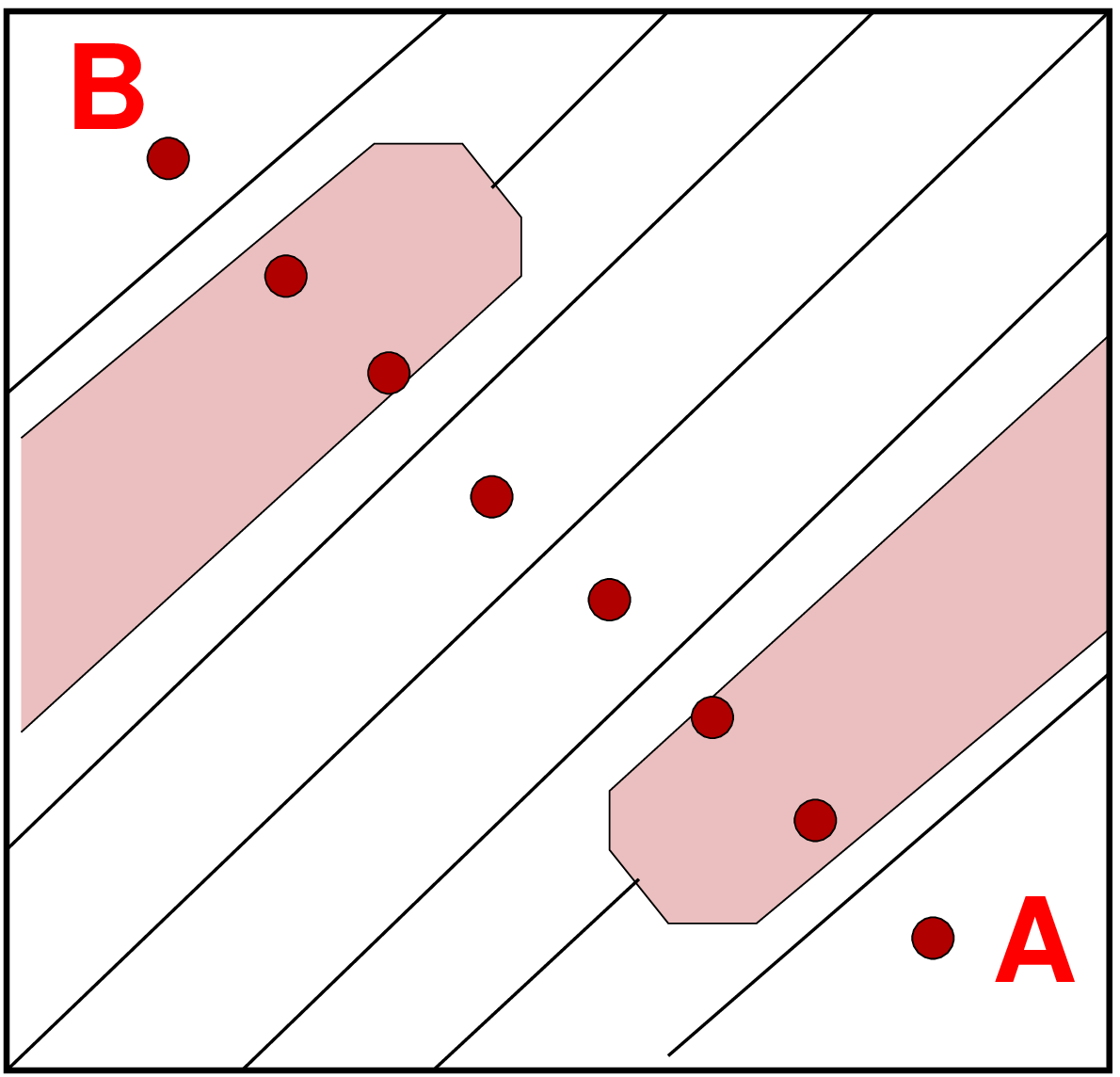}}}}\hspace{1cm}\makebox[20pt][l]{(b)}{\rotatebox{0}{{\includegraphics[scale=0.4,clip=true]{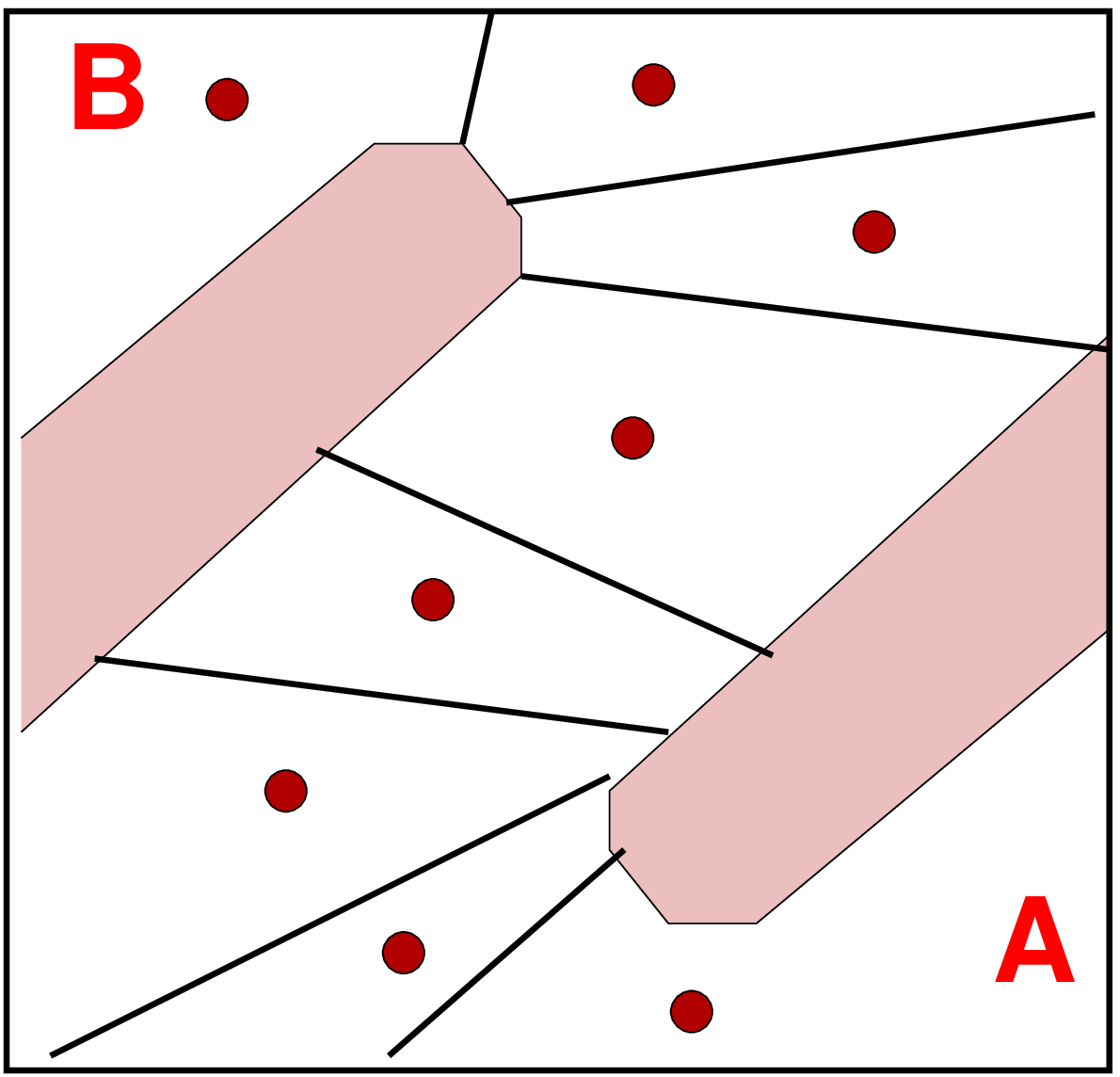}}}}\caption{A two-dimensional potential energy landscape for which the reaction coordinate is Z-shaped and cannot be described by a simple linear combination of the x and y coordinates of the system. (a): Partitioning of space with Voronoi polyhedra around a linear string of beads between the A and B states; this gives interfaces which follow the reaction coordinate poorly. (b): Partitioning into interfaces using a curved string of beads; this gives a much better set of interfaces.\label{fig:zorro} }
\end{center}
\end{figure}

For some rare event problems, it is easy to define a good order parameter. For example, for crystal nucleation processes, one typically chooses the number of particles in the system that are ``crystalline''  \cite{pieterrein}, for a polymer translocation problem one can use the number of translocated monomers  \cite{FFS2}, and for a bistable chemical reaction in which the transition is between states rich in molecules of chemical species A and B, one can use the difference between the number of A and B molecules \cite{FFS,FFS2,morelli_jcp,morelli_bpj,morelli_lambda}. However, in other cases the choice of order parameter is less obvious. For example, for hydrophobic polymer collapse the solvent coordinates as well as those of the monomers can play an important role in the transition \cite{pr_chandler}. Complex reaction coordinates are also a common feature of protein folding problems \cite{bolhuis_pnas}. Figure \ref{fig:zorro} illustrates a simple two-dimensional potential energy landscape for which the Z-shaped reaction coordinate cannot be described by either of the two coordinates of the system (x and y), or by any linear combination of these coordinates \cite{dickson2009}.

For complex cases such as the one illustrated in Figure \ref{fig:zorro}, the Voronoi tessellation approach pioneered by Vanden-Eijnden and Venturoli \cite{Venturoli2} can prove very useful.  Here, one defines a path connecting A and B by a set of configurations, or ``beads''. The path need not be describable in terms of a single collective coordinate. Any configuration of the system can be classified according to which of the beads it lies closest to: this is equivalent to partitioning space into a set of Voronoi polyhedra around the beads. Interfaces can then be defined as the planes in phase space across which the ``closest bead identity'' changes, or equivalently the edges of the Voronoi polyhedra. Figure \ref{fig:zorro} shows the interfaces defined by this method for our two-dimensional example, for a linear set of beads [panel (a)] and for a set of beads chosen to lie along the curved reaction coordinate [panel(b)] \cite{anderson}: it is clear that Voronoi tessellation provides a very convenient and easy-to-implement way to translate a string of beads into a set of interfaces. It is important to note, however, that for highly multidimensional problems it is likely to be necessary to project the phase space onto a small number of order parameters before carrying out the Voronoi tessellation, otherwise the resulting interfaces may be highly convoluted. This approach does not therefore entirely eliminate the need for order parameters. In addition, of course, one still needs to find a suitable set of beads, for example by an iterative technique such as the finite temperature string method (FTS) \cite{e_05,Venturoli2}. In Section \ref{sec:neus}, we discuss briefly how this is done in the context of nonequilibrium Umbrella sampling \cite{dickson2009}.

\subsection{The committor and the reaction coordinate}
The  committor function $P_B(x)$ is defined as the probability that a trajectory initiated from configuration $x$ will reach the final state B before the initial state A.  Along a transition path, the committor function increases from zero to unity.  Configurations along the transition paths for which $P_B=0.5$ have special significance: the collection of these configurations is known as the ``transition state ensemble'', or TSE (although alternative definitions are also possible \cite{hummer2004,best}). Analysis of the TSE configurations can provide insight into the reaction mechanism. If the probability distribution for a given order parameter, evaluated over the TSE configurations, is highly peaked, it is likely that this order parameter closely corresponds to the reaction coordinate, whereas a broad or bimodal distribution indicates that other order parameters are needed to fully describe the transition mechanism \cite{bolhuis_arpc,dellago2}. Scatter plots for the TSE configurations as functions of various collective coordinates can also provide insight into the reaction mechanism, as discussed in Section \ref{sec:apps}.  To extract committor values for configurations along the transition paths, one typically fires a large number of trial runs from each configuration to estimate the probability of these reaching B rather than A. This is a computationally expensive procedure (although some effort can be saved if one is only interested in the TSE). In recent work, however, Borrero and Escobedo \cite{borrero2007}  have shown that committor values can be extracted on-the-fly from FFS simulations, by making intelligent use of the information already obtained on the number of successful trials to interface $\lambda_{i+1}$ for each configuration at interface $\lambda_i$. This  is discussed in Section \ref{sec:lse}.

The committor function $P_B$ is in some sense the ideal reaction co-ordinate, since it by definition correlates with the progress of the transition. However,  $P_B$ is a complex function of all the coordinates of the system. To obtain scientific insight, one needs to be able to project this function onto a small set of physically meaningful collective coordinates. Hummer {\em{et al}} \cite{hummer2004,best} proposed a variational method for optimising reaction coordinates, based on evaluating the projection onto the order parameter of the probability function $p(TP|x)$ that a configuration $x$ forms part of a transition path. Ma and Dinner \cite{ma} proposed a method in which one uses ``representative'' configurations corresponding to different values of the committor $P_B$ to determine the functional dependence of the committor on each of a chosen set of collective coordinates. An optimisation procedure (in this case a genetic algorithm) can then be used to find the best combination of these coordinates to represent the committor.  In related work, Peters {\em {et al}} \cite{peters1,peters2} also proposed a method for determining the optimum combination of collective coordinates to represent the committor. In their method, the committor values are obtained on-the-fly using a version of transition path sampling called Aimless Shooting. A simple model for the reaction coordinate (eg a linear combination of collective coordinates) is assumed and the parameters of the model optimised using Bayesian likelihood maximisation.

\subsection{Using the committor to optimise the order parameter in FFS}\label{sec:lse}
\begin{figure}[h!]
\begin{center}
{\rotatebox{0}{{\includegraphics[scale=0.55,clip=true]{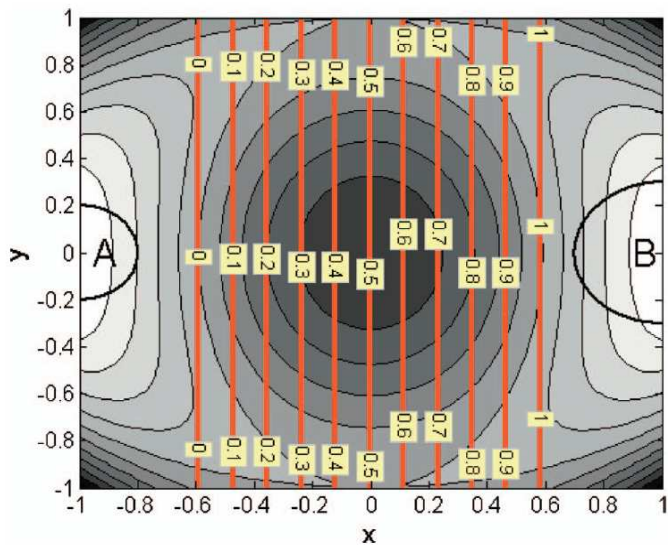}}}}
\caption{Isocommittor lines obtained using FFS-LSE for a simulation on the two dimensional potential energy landscape represented by the contour plot. The numbers indicate the committor values. [Reproduced with kind permission from Ref \cite{borrero2007}].\label{fig:borrero1} }
\end{center}
\end{figure}

In recent work, Borrero and Escobedo have proposed a method (related to that of Peters {\em{et al}} \cite{peters1,peters2}) in which information on the committor is extracted directly from FFS simulations, and used to optimise the choice of order parameter \cite{borrero2007}. This method is known as FFS-LSE (least square estimation). Borrero and Escobedo show that for the branched growth variant of FFS (see Section \ref{sec:variants}), the committor value $P_{Bj}^i$ for configuration $j$ at interface $\lambda_i$ is given by:
\begin{equation}\label{eq:comm}
P_{Bj}^i = p_j^i(\lambda_{i+1}|\lambda_i)\frac{\sum_{m=1}^{N_j^i}P_{Bm}^{i+1}}{N_j^i} = \frac{\sum_{m=1}^{N_j^i}P_{Bm}^{i+1}}{k_i}
\end{equation}
where $k_i$ is the number of trials per configuration at interface $i$, $N_j^i$ is the number of successful trials from configuration $j$ at interface $i$ and $p_j^i(\lambda_{i+1}|\lambda_i) = N_j^i / k_i$ is the probability of success for configuration $j$ at interface $i$. This equation states that the committor value for configuration $j$ is the probability that a trial run fired from this configuration will reach the next interface $\lambda_{i+1}$, multiplied by the average committor value for its ``daughter'' configurations at $\lambda_{i+1}$. If information on the connectivity between configurations at successive interfaces, as well as on the number of successful trials for each configuration, is stored during an FFS run, committor values for these configurations can be extracted at the end of the run with no additional computational effort. Although Eq.(\ref{eq:comm}) was derived for the BG algorithm, it should be possible to apply a similar approach to the other FFS variants.

This procedure produces a set of configurations with associated committor values over the whole range of the order parameter $\lambda$. For these configurations, one evaluates a set of  $m$ candidate collective coordinates $q$. This data is then used to fit a parametrised functional form for the dependence of the committor on the  $q$:
\begin{equation}
P_B(q) = \sum_{k=1}^m \beta_k q_k + q^T {\bf{A}} q + \beta_0 + \epsilon
\end{equation}
where the $\beta$ values and the matrix ${\bf{A}}$ are optimised by least squares fitting to the data set obtained from FFS, and $\epsilon$ is the sum of the squares of the errors, to be minimised. The resulting functional form for $P_B(q)$ is the optimal choice for the FFS order parameter $\lambda(q)$. Choosing the order parameter $\lambda(q)$ that most closely matches the committor function  (within the constraints of the fitting function) should lead to interfaces that are perpendicular to the transition paths, resulting in an efficient computation of the rate constant.  Borrero and Escobedo found that a single iteration was sufficient to converge on the optimal $\lambda(q)$, for several test cases including the flipping of a model genetic switch and the folding of a lattice protein model \cite{borrero2007}. Figure \ref{fig:borrero1} shows the isocommittor lines obtained using this procedure, for a test simulation on the two dimensional potential surface represented by the contour plot. It might be interesting in future to  combine this approach with a bead-string description of the order parameter as discussed in Section \ref{sec:voronoi}.

\section{Computing stationary distributions}\label{sec:sd}
For many rare event problems, one is interested not only in the rate constant and transition paths, but also in the steady-state probability distribution $\rho$, as a function of one or more order parameters $q$. Knowledge of the steady-state distribution allows one to compute, for example, averages of experimentally measurable observables for comparison with experiments. For systems with stable A and B states, $-\ln{\rho(q)}$ takes the form of a ``barrier'' with a peak separating the two stable states. If the dynamics of the system obeys detailed balance, this distribution is directly related to the free energy function $F$: $F(q) \sim -\ln{\rho(q)}$.   For these equilibrium systems, $\rho(q)$ can be computed by umbrella sampling \cite{daan}, in which one divides the range of $q$ into a series of windows, runs a separate simulation in each window and uses the Boltzmann distribution to reassemble the probability distributions from each window into the unbiased $\rho(q)$. In this section, we first discuss how  $\rho(q)$ can be extracted from FFS simulations, for equilibrium or nonequilibrium systems \cite{barrier}, and a related method for computing $\rho(q)$ with FFS, for equilibrium systems only \cite{borrero2009}. We also briefly discuss a method which allows $\rho(q)$ to be computed for nonequilibrium systems in multiple dimensions \cite{warmflash2007,dickson2009}.

We note that for these computations there is no requirement for a stable B state. In addition to computing barrier heights, these methods can also be used  to explore the probability distribution for excursions of the system from a stable A state, along an order parameter, without the presence of a stable B state. In this case, $-\ln{\rho(q)}$ will be an increasing function, whose detailed shape contains information on the physics of fluctuations away from the A state. 

\subsection{Obtaining stationary distributions from FFS simulations}\label{sec:barrier}
The stationary distribution function $\rho(q)$ can be obtained on-the-fly during an FFS simulation, in a similar approach to that of Moroni {\em{et al}} for computing $\rho(q)$ in PPTIS simulations \cite{Moroni3}. To achieve this, one computes  histograms for the $q$ values visited by all trial runs (failed and successful), fired from interface $i$. In the case of a stable B state, it is necessary to run the FFS calculation in both directions (A$\to$B and B$\to$A) \cite{barrier}. 

\begin{figure}[h!]
\begin{center}
{\rotatebox{0}{{\includegraphics[scale=0.35,clip=true]{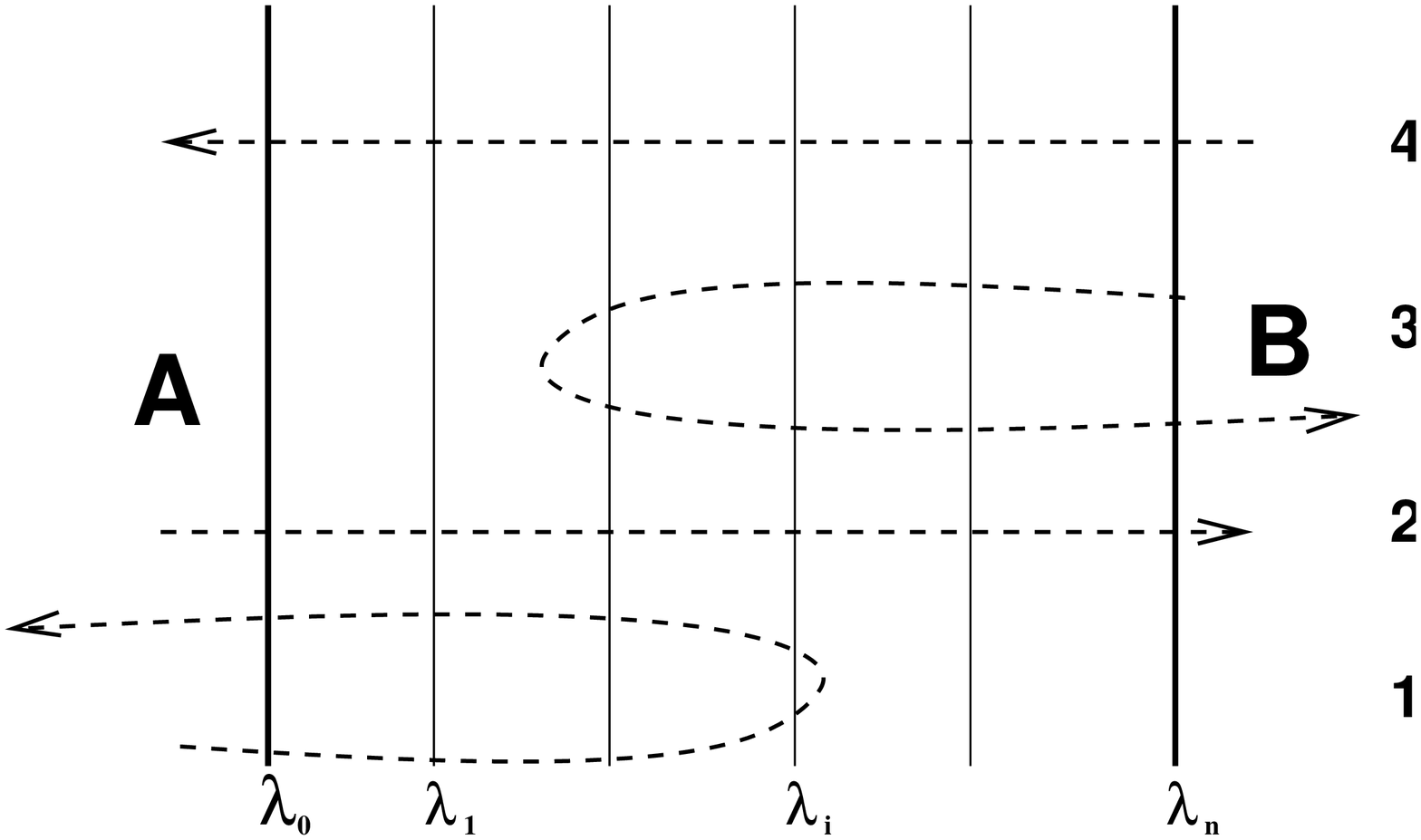}}}}
\caption{Illustration of categories of trajectories contributing to the stationary distribution $\rho(q)$. Trajectories 1 and 2 originate in A and are sampled by an FFS simulation from A to B. Trajectories 3 and 4 originate in B and are sampled by an FFS simulation from B to A.   [Reproduced with permission from Ref \cite{barrier}].\label{fig:sketch} }
\end{center}
\end{figure}

The principle underlying the calculation of $\rho(q)$ with FFS is illustrated in Figure \ref{fig:sketch}. Trajectories visiting a particular region of phase space can be grouped according to their origin in either the A or the B state.  The stationary distribution function can be written as the sum of the contributions, $\psi_A$ and $\psi_B$ respectively, of these two groups of trajectories:
\begin{equation}\label{eq:rho}
\rho(q)=\psi_A(q) + \psi_B(q)
\end{equation}
where $\psi_A$ and $\psi_B$ can be written as:
\begin{equation}
\psi_A = \rho_A \Phi_A \tau_+(q;\lambda_0) \qquad \qquad \psi_B = \rho_B \Phi_B \tau_-(q;\lambda_n)
\end{equation}
Here, $\rho_A$ is the steady-state probability of finding the system in A, $\Phi_A$ is the flux of trajectories out of the A state and $\tau_+(q;\lambda_0)$ is the average time spent with a value $q$ of the order parameter, for a trajectory which originates from $\lambda_0$. The equivalent definitions hold for  $\rho_B$ and $\Phi_B$, while  $\tau_-(q;\lambda_n)$ is the equivalent average time, for a trajectory which originates from $\lambda_n$. These averages must be taken over all trajectories leaving $\lambda_0$ (or $\lambda_n$), whether or not they eventually reach B (or A), with the correct statistical weights. 

In an FFS simulation from A to B, $\Phi_A$ is computed during the initial simulation in the A basin. The probabilities $\rho_A$ and $\rho_B$ can be obtained once the rate constants $k_{AB}$ and $k_{BA}$ have been computed, since in steady state $\rho_A k_{AB} = \rho_B k_{BA}$ and $\rho_A + \rho_B=1$ (assuming transitions are fast compared to the time spent in the A and B basins). The function $\tau_+(q;\lambda_0)$  can be obtained if we measure during the FFS simulation the average time $\pi_+(q;\lambda_i)$ spent with order parameter value $q$, for a trial run fired from $\lambda_i$. This average should be computed over all configurations in all trial runs fired from $\lambda_i$, regardless of whether they succeed in reaching $\lambda_{i+1}$, and should include any differential weighting factor applied to the trial runs (e.g. in the case of the RB method or if pruning is used). $\tau_+(q;\lambda_0)$ is then given by:
\begin{equation}\label{eq:tau}
\tau_+(q;\lambda_0) = \pi_+(q;\lambda_0) + \sum_{i=1}^{n-1}\pi_+(q;\lambda_i)\prod_{j=0}^{i-1}P(\lambda_{j+1}|\lambda_j)
\end{equation}
Eq.(\ref{eq:tau}) expresses the average time $\tau_+(q;\lambda_0)$ spent at order parameter value $q$ as the sum of contributions from partial trajectories (trial runs) originating at each interface, weighted by the probability $\prod_{j=0}^{i-1}P(\lambda_{j+1}|\lambda_j)$ of observing such a partial trajectory in a brute-force simulation. As the interface index $i$ increases, this probability decreases, but $\pi_+(q;\lambda_i)$ continues to be well-sampled. This is because FFS allows good sampling of regions of the phase space which are rarely visited in a brute-force simulation. A similar procedure can be applied in the B to A direction to obtain $\Phi_B$, $k_{AB}$ and $\tau_-(q;\lambda_n)$.  A detailed description of the practical implementation is given in Ref \cite{barrier}. We note that if the B state is very much less stable than the A state (or not stable at all), one can make the approximation $\rho_B \approx 0$, so that $\psi_B \approx 0$ in Eq.(\ref{eq:rho}) and only a simulation in the A to B direction is required. It is also important to point out that  $q$ need not correspond to the FFS order parameter $\lambda$: one may even obtain the distribution as a function of multiple coordinates $\rho(q_1,q_2)$, simply by computing multidimensional histograms $\pi_\pm(q_1,q_2;\lambda_i)$ during the FFS simulation. However, for order parameters very different from $\lambda$, the advantage of enhanced sampling of rarely visited regions of the phase space will be lost.

\begin{figure}[h!]
\begin{center}
{\rotatebox{0}{{\includegraphics[scale=0.35,clip=true]{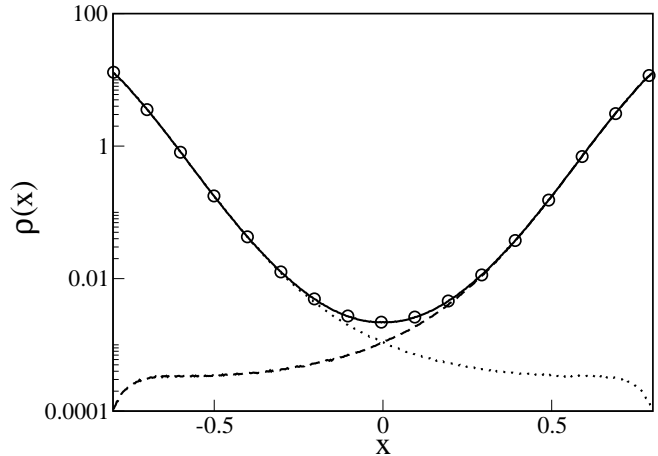}}}}
\caption{Stationary distribution $\rho(x)$ (solid line) obtained using FFS, compared to the 
normalised Boltzmann distribution (circles) for a symmetric double well potential $V(x) = ax -bx^2 + cx^4$ with $a=0.25$, $b=2$, $c=1$, simulated using overdamped Brownian dynamics with diffusion constant $D = 0.01$, $k_BT=0.1$ and $dt=0.05$. The dotted and dashed lines 
show $\psi_A(x)$ and $\psi_B(x)$ respectively [Reproduced with permission from Ref \cite{barrier}].\label{fig:barrier} }
\end{center}
\end{figure}

Figure \ref{fig:barrier} shows $\rho(x)$ for a simple one-dimensional potential, computed using FFS and compared with the expected Boltzmann distribution. Similarly convincing results were obtained for the nucleation barrier in a  two-dimensional Ising model \cite{barrier}, and for the nonequilibrium case  of the flipping of a model genetic switch \cite{barrier,morelli_bpj}.

\subsection{Forward flux / umbrella sampling}
The method described in Section \ref{sec:barrier} requires FFS simulations in both the forward and backward directions, if the steady-state population of the B state is significant. An approach developed recently by Borrero and Escobedo, known as Forward Flux / Umbrella Sampling (FFS-US) \cite{borrero2009}, aims to avoid this requirement, for systems where detailed balance is obeyed. In FFS-US, histograms obtained in a FFS simulation in the direction A$\to$B are combined with histograms obtained with conventional umbrella sampling \cite{daan}, using the interfaces as hard walls. For the umbrella sampling, trajectories are initiated inside each window using configurations obtained in the FFS simulation. The umbrella sampling histograms correct for the bias that arises in the FFS histograms due to the fact that one simulates only in the A$\to$B direction.

\subsection{Nonequilibrium umbrella sampling in multiple dimensions}\label{sec:neus}
A method has recently been proposed by Warmflash {\em{et al}} \cite{warmflash2007}, and extended by Dickson {\em{et al}} \cite{dickson2009}, which allows for efficient computation of multidimensional steady state distributions for nonequilibrium rare event problems.  Although this approach is distinct from FFS, we feel it is useful to include a brief description of it in this review, since it is one of the few methods that give access to the steady state distribution for nonequilibrium systems.

 In the method proposed by Warmflash {\em{et al}} \cite{warmflash2007},  the phase space region of interest is divided into a lattice  using one or several order parameters. Separate simulations are run in each lattice box simultaneously, during which one counts the numbers of simulation trajectories which attempt to transfer between boxes. When this happens, ``weight'' is transferred between boxes, and the trajectory is reinserted at the boundary of the same box with a configuration drawn from a self-consistently determined statistical distribution. To achieve this, one simulates at the same time a second lattice of boxes, with a grid offset relative to the first lattice. The second lattice provides configurations corresponding to interface crossings that can be used in the first lattice, and {\em{vice versa}} [for details see Refs \cite{warmflash2007} and \cite{dickson2009}]. The stationary distribution is finally obtained from the simulated probability distributions within each box, multiplied by the weight computed for that box. This method, unlike FFS, is suitable for problems with slow dynamics in multiple dimensions. However, for many dimensions its implementation is likely to become complicated. The more recent work by Dickson {\em{et al}}  \cite{dickson2009} makes several modifications. Here, the order parameter instead consists of a string of beads, with the boxes now being defined by Voronoi polyhedra (see Section \ref{sec:voronoi}).  A similar approach  is used, except that one now simulates two strings, each offset from the other, and the algorithm for reinserting trajectories is somewhat different. Importantly, however, one can now use the computed distributions at the interfaces to iteratively improve the positioning of the beads, in an approach inspired by the string method \cite{Ren02,ren2005}. Although Refs \cite{warmflash2007} and \cite{dickson2009} focus on the stationary distribution function $\rho(q)$, it seems likely that this approach could also give access to rate constants, in a similar manner to the Weighted Ensemble method of Huber and Kim \cite{huber1996}. 

\section{Applications}\label{sec:apps}

Forward flux sampling has been applied to quite a number of different equilibrium and nonequilibrium rare event problems, with a variety of simulation techniques including Metropolis Monte Carlo, Molecular Dynamics, Brownian Dynamics and kinetic Monte Carlo. Transitions studied include nucleation in a variety of different contexts  \cite{ising_shear1,ising_shear2,page2006,sear2006,sear2007,sanders2007,valeriani2007,valeriani_carbon,valeriani2005,sanz2007,sanz2008,barrier,koos_amanda,wang}, genetic switch flipping \cite{FFS,morelli_jcp,morelli_bpj,morelli_lambda}, changes in DNA configuration \cite{moebius2006}, droplet coalescence \cite{rekvig2007}, polymer translocation \cite{FFS2,huang2008,hernandez2008} and protein conformational changes \cite{borrero2006,vega}. It is not our purpose here to  discuss these applications in detail (an excellent review of biomolecular applications is given in \cite{escobedo}). Instead, we present a brief overview of three applications with which we have been involved, highlighting the contributions made using FFS as well as particular methodological challenges.

\subsection{Genetic switch flipping}\label{sec:switch}

\begin{figure}[h!]
\begin{center}
\makebox[20pt][l]{(a)}{\rotatebox{0}{\includegraphics[scale=0.55,clip=true]{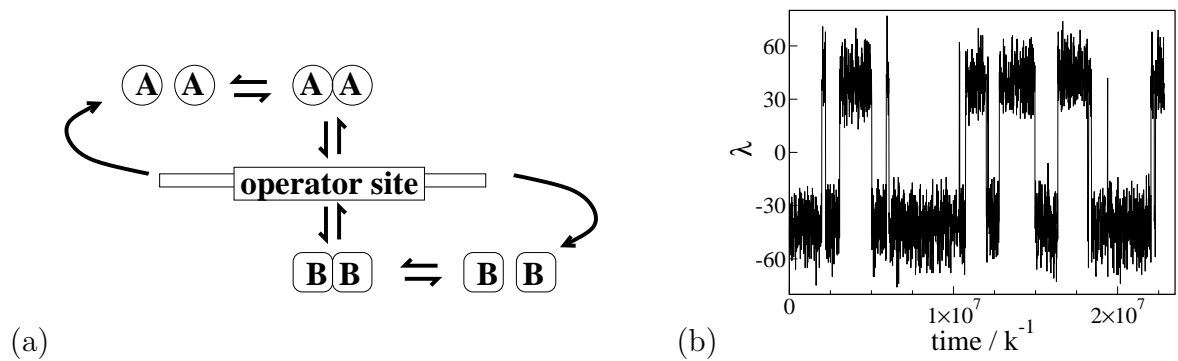}}}\hspace{1cm}\makebox[20pt][l]{(b)}{\includegraphics[scale=0.25,clip=true]{allen_etal_fig12b.eps}}
\caption{(a) Schematic illustration of the model genetic switch (b) A typical brute-force simulation trajectory. The order parameter $\lambda \equiv N_{A}-N_{B}$ is plotted as a function of time, where $N_{A}$ and $N_{B}$ are the total number of A and B molecules respectively.  [Reproduced with permission from Ref \cite{FFS2}].\label{fig:switch1} }
\end{center}
\end{figure}

\begin{table}[h]
\begin{center}
\begin{tabular}{cccc|cc}
\multicolumn{3}{c}{Reactions} &\,\, & \multicolumn{1}{c}{Forward rate constant}&\multicolumn{1}{c}{Backward rate constant}\\
$2{\mathrm{A}} \rightleftharpoons {\mathrm{A}}_2$ &\,\,& $2{\mathrm{B}} \rightleftharpoons {\mathrm{B}}_2$ &\,\,&\, $5k$ & $5k$\\ 
${\mathrm{O}} + {\mathrm{A}}_2 \rightleftharpoons {\mathrm{O}}{\mathrm{A}}_2$ &\,\,& $ {\mathrm{O}} + {\mathrm{B}}_2 \rightleftharpoons {\mathrm{O}}{\mathrm{B}}_2$ &\,\,&\, $5k$ & $k$\\
$ {\mathrm{O}} \to {\mathrm{O}} + {\mathrm{A}}$ &\,\,& $ {\mathrm{O}} \to {\mathrm{O}} + {\mathrm{B}}$ &\,\,& \,$k$ & -\\
$ {\mathrm{O}}{\mathrm{A}}_2 \to {\mathrm{O}}{\mathrm{A}}_2 + {\mathrm{A}}$ &\,\,& ${\mathrm{O}}{\mathrm{B}}_2 \to {\mathrm{O}}{\mathrm{B}}_2 + {\mathrm{B}}$ & \,\,&\,$k$ & -\\
$ {\mathrm{A}} \to \emptyset$ &\,\,& ${\mathrm{B}} \to \emptyset$ &\,\,&\, $0.25k$ & -\\\\
\end{tabular}
\caption{Reaction scheme for the model genetic switch. The unit of time is $k^{-1}$.\label{tab:switch}}
\end{center}
\end{table}

Gene regulatory networks control the behaviour of biological cells. In these  networks, genes encode protein molecules which in turn  control the expression of other genes. The resulting networks of interactions between genes allow cells to perform the computations necessary for survival and proliferation. Of particular interest are gene regulatory networks with multiple stable states, corresponding to alternative cellular developmental outcomes. A classical example is the bistable bacteriophage $\lambda$ switch, which controls the transition between lysogeny (quiescent integration into the host cell) and lysis (replication and killing of the host cell) for phage $\lambda$, a virus which infects the bacterium {\em{Escherichia coli}} \cite{Ptashne86}. We studied a highly simplified representation of the gene network controlling this switch. In this model switch, two genes $A$ and $B$  mutually repress
one other (see Figure \ref{fig:switch1}a). If gene $A$ is turned on, protein A is
produced. This protein can dimerise and in the dimer form it can bind to the DNA and prevent the production of protein B from gene $B$. There is thus a stable state with a high concentration of protein A and a low concentration of protein B.  If, however, due to a
fluctuation, protein A dissociates from the DNA, then gene $B$ can be
expressed. The newly synthesised protein B can dimerise and bind to the
DNA, preventing the expression of gene $A$. This can ultimately lead to the flipping of the switch into the alternative stable state with a high concentration of protein B.  Figure \ref{fig:switch1}b shows a typical trajectory for a brute-force simulation of this model switch. The system undergoes infrequent but rapid  flips between the A-rich and B-rich states. These random flips are driven by intrinsic biochemical ``noise'': fluctuations due to the stochasticity of chemical reactions. Recent experiments have shown that this noise can play an important role in gene regulation \cite{raj}; yet it is known that the bacteriophage $\lambda$ genetic switch is extremely stable, with a spontaneous flipping rate of less than once in every $10^{9}$ generations \cite{Little99}. This raises the question of what the mechanisms are that govern that stability of bistable genetic switches in the presence of biochemical noise. We have addressed this question using FFS.

We were interested how the detailed rules for DNA binding affect the switch stability: in particular, the comparison between a switch in which a dimer of protein A excludes the binding of B dimers to the DNA (and vice versa), with a switch in which both dimers can bind simultaneously \cite{Warren04,Warren05,FFS}. The former is termed an ``exclusive switch'' and the latter a ``general switch''. The exclusive switch is described by the set of chemical reactions given in Figure \ref{fig:switch1}a. In this reaction scheme, $O$ represents a DNA regulatory sequence (operator)
adjacent to two divergently transcribed genes $A$ and $B$, which code
respectively for proteins A and B. These can dimerise to form ${\rm{A_2}}$ and   ${\rm{B_2}}$. Genes
$A$ and $B$ can each produce proteins with the same rate, but
whether they do so depends on the state of the operator $O$.  When
an ${\rm{A_2}}$ dimer is bound to $O$, the production of B is blocked,
and likewise, when a ${\rm{B_2}}$ dimer is bound to $O$, the
production of A is blocked. Proteins can also vanish (in
the monomer form), modelling degradation and dilution in a cell. The reaction set for the general switch is identical except for the addition of the extra reactions $O{\rm{A_2}} + {\rm{B_2}}\rightleftharpoons O{\rm{A_2}}{\rm{B_2}}$ and $O{\rm{B_2}} + {\rm{A_2}}\rightleftharpoons O{\rm{A_2}}{\rm{B_2}}$ ({\em{i.e.}} the operator can bind both dimers simultaneously): when dimers are bound to the
operator, neither protein can be produced.

\begin{figure}[h!]
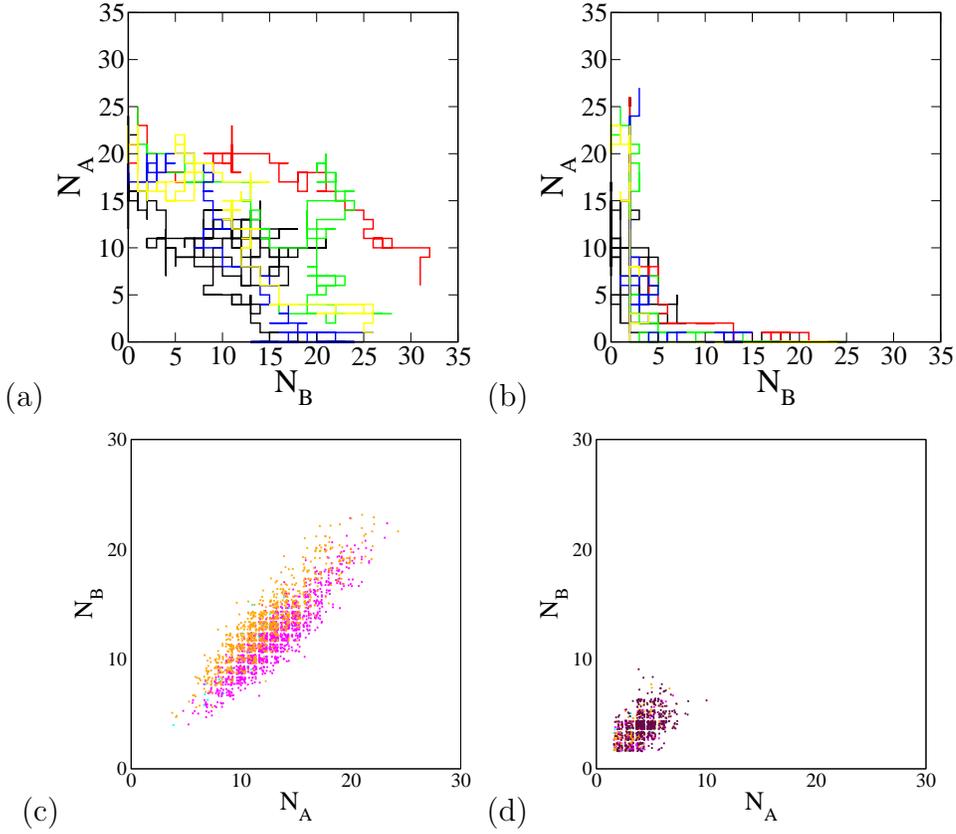

\begin{center}
\makebox[20pt][l]{(a)}{\rotatebox{0}{\includegraphics[scale=0.29,clip=true]{allen_etal_fig13a.eps}}}\hspace{0.2cm}\makebox[20pt][l]{(b)}{\includegraphics[scale=0.29,clip=true]{allen_etal_fig13b.eps}}\\\vspace{0.2cm}\makebox[20pt][l]{(c)}{\rotatebox{0}{\includegraphics[scale=0.29,clip=true]{allen_etal_fig13c.eps}}}\hspace{0.2cm}\makebox[20pt][l]{(d)}{\includegraphics[scale=0.29,clip=true]{allen_etal_fig13d.eps}}
\caption{Top: Five randomly chosen transition paths, plotted in the $N_{A}-N_{B}$ plane, for  (a) the exclusive and (b) the general switch. Each transition path is shown in a different colour. Bottom: TSE configurations plotted in the $N_{A}-N_{B}$ plane, and colour coded according to operator state, for (c) the exclusive switch and (d) the general switch. Cyan: $O$, gold: $O{\rm{A_2}}$, magenta:  $O{\rm{B_2}}$, violet: $O{\rm{A_2 B_2}}$.  [Panel (a) reproduced with permission from Ref \cite{FFS}].\label{fig:switch2} }
\end{center}
\end{figure}

We simulated this model using brute force simulation and DFFS, using the order parameter $\lambda \equiv N_{A}-N_{B}$, where $N_{A}$ and $N_{B}$ are the total number of A and B molecules respectively. Our results showed that the rules for operator binding can have  a
dramatic effect on the stability of the switch: for typical parameter values, the exclusive switch is
orders of magnitude more stable than the general switch
\cite{Warren04,Warren05,FFS}. To understand this result, we extracted transition paths from our FFS simulations. Figures \ref{fig:switch2}a and \ref{fig:switch2}b show typical
switching pathways for the general and exclusive switch,
plotted in the $N_{A}-N_{B}$ plane \cite{FFS}. The switching pathways are very different: during a typical
flipping trajectory, the general switch passes through a region where
the copy numbers $N_A$ and $N_B$ are both nearly zero, in
contrast to the exclusive switch. To characterise the switching
pathways further, we extracted configurations from the transition state ensemble (TSE). These configurations are plotted in the $N_{A}-N_{B}$ plane in Figures \ref{fig:switch2}c and \ref{fig:switch2}d, where the colour coding illustrates the operator state. These plots show that for the general switch, the  TSE is dominated
by configurations in which both proteins are bound to the operator and the production of both genes is repressed, which explains why
at the top of the ``barrier'', $N_A$ and $N_B$ are both nearly
zero. In the exclusive switch, however, these states are not allowed (since the proteins mutually exclude each other's binding). Thus, in the transition paths,  $N_A$ and
$N_B$ never become nearly zero simultaneously. This explains why the exclusive switch is more stable than the general switch. Due
to rare fluctuations, copies of the minority species will
occasionally be produced. In the general switch, these can immediately bind the operator,
leading to the repression of the majority species, which is the
critical step that ultimately leads to the flipping of the switch.  In
the exclusive switch, however, a newly synthesised dimer of the
minority species probably cannot bind the operator, since it is likely to be blocked by binding of the majority species. The system then has to wait for another fluctuation whereby the majority species dissociates from the operator before a
flipping event can be initiated.

The model described above is a highly simplified representation of a real genetic switch.  Recently, we
have simulated a much more detailed model of the bacteriophage $\lambda$ switch
\cite{morelli_lambda}. This system comprised over 500 chemical reactions, which were computationally expensive to simulate, requiring the use of FFS as well as  dynamical coarse-graining of some of the chemical equilibria (although importantly not of the operator binding reactions). Again using DFFS with the same choice of order parameter as for the simple model switch, we were able to reproduce the extreme stability of the switch, as observed experimentally. These simulations also revealed a key role for a DNA looping interaction (not present in the simple model), in maintaining the extraordinary stability of the switch.

These results show that rare event simulation methods
such as FFS can successfully be applied to study rare events in
complex biochemical networks whose dynamics is intrinsically out of equilibrium. Multistability has been found to play an important role in many different biological contexts, ranging
from cell differentiation, apoptosis, the immune system, to the cell
cycle. We hope that FFS will prove a useful tool for modelling these processes.

\subsection{Homogeneous crystal/bubble nucleation}

Phase transitions occurring by homogeneous nucleation are a widespread and important class of rare event processes. When a liquid is cooled below its melting point, the liquid state becomes metastable with respect to the crystal. The supercooled system can spend a long time in the liquid state before undergoing a rapid transformation into the thermodynamically stable solid phase. Similarly, a liquid that is heated above its boiling point can undergo a ``cavitation'' transition in which bubbles of the gaseous phase are formed. Since these nucleation processes are rare events, brute-force simulations are often impractical. However, rare event methods such as FFS can be used to calculate rate constants and
transition pathways. Nucleation in ``quasi-equilibrium'' systems whose dynamics obeys detailed balance has been tackled using a range of different rare event simulation methods, including PPTIS \cite{Moroni4},
Metadynamics \cite{parrLJ}
committor analysis combined with two-dimensional Umbrella sampling
scheme~\cite{radha},  Mean First Passage Time calculations~\cite{reguera2,reguera3} and  order-parameter-based Monte Carlo simulation~\cite{pablo}. We refer to these as ``quasi-equilibrium'' rather than ``equilibrium'' systems because for nucleation problems the initial state is always metastable with respect to the final state. For such systems, FFS provides a complementary approach to these methods. FFS also provides the potential  for studying   nucleation phenomena in out-of-equilibrium systems whose dynamics does not obey detailed balance (for example, with applied external shear, as discussed in the following section). 

Over the last few years, FFS has been used to study crystal nucleation 
in both covalent \cite{valeriani2007,valeriani_carbon} 
and ionic quasi-equilibrium systems \cite{valeriani2005,sanz2007,sanz2008}. For these systems, it is possible to compare the results of FFS to those obtained by other rare event methods. In particular, in Ref.\cite{valeriani2005}, we studied the nucleation of crystalline sodium chloride from the melt at  moderate super-cooling. The nucleation rate was computed with a Bennett-Chandler procedure, in which Umbrella Sampling was used to compute the free energy barrier, followed by the firing of trajectories from the top of the barrier to obtain the transmission coefficient. We also computed the nucleation rate using FFS. In both cases, the order parameter was taken to be the size of the largest solid cluster.  Both methods yielded 
the same nucleation rate, to within the statistical error bars. In later work, \cite{barrier}, we used FFS to compute the nucleation free energy barrier for a  two-dimensional Ising system, and compared the results to those of Umbrella Sampling. Both methods gave (to within error bars) the same free energy barrier height and shape. FFS has also recently been used to study vapour-crystal nucleation \cite{koos_amanda}.  

Recently, we combined FFS with Molecular Dynamics simulations to study  bubble nucleation (cavitation) \cite{wang}, and obtained results that differed from those of Umbrella Sampling. We computed the nucleation rate using FFS and also analysed the  transition path ensemble. Our results showed that cavitation starts with compact bubbles rather 
than ramified ones, as has previously been suggested by Umbrella Sampling \cite{debenedetti}. The FFS method does allow for the  formation of ramified structures, but 
these pathways are kinetically unfavourable. Such kinetic effects cannot be  observed in the Umbrella 
Sampling scheme \cite{debenedetti}, which assumes a local thermal equilibrium. Our FFS simulations also indicated a strong correlation between local temperature 
fluctuations in the liquid and subsequent bubble formation. Kinetic effects in homogeneous nucleation have also been investigated in the context of the crystallisation of a  binary mixture of oppositely charged colloids, interacting via Yukawa potentials \cite{sanz2007,sanz2008}. Here, FFS simulations suggested that, for some thermodynamic conditions, the growth of the crystal 
does not follow the minimum free-energy path. This indicates a lack of ergodicity 
of the fluid on the timescale of crystal growth. This effect could not have been observed with a technique such as Umbrella Sampling, which assumes quasi-equilibrium transition paths. 

These studies constitute a promising start for FFS in shedding new light on nucleation processes. For systems where the nucleation paths can be expected to be in local thermal equilibrium, FFS results agree well with those of other methods (such as Umbrella Sampling), whereas for several cases where kinetic effects are important, FFS has revealed unexpected behaviour. Plenty of scope remains for further investigation of such effects.

\subsection{Nucleation in a sheared Ising model}
\begin{figure}[h!]
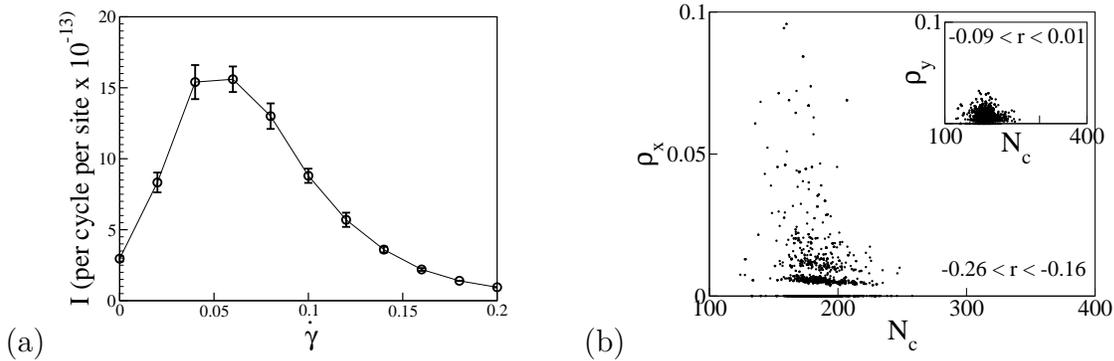

\begin{center}
\makebox[20pt][l]{(a)}{\rotatebox{0}{{\includegraphics[scale=0.25,clip=true]{allen_etal_fig14a.eps}}}}\hspace{1cm}\makebox[20pt][l]{(b)}{\rotatebox{0}{{\includegraphics[scale=0.25,clip=true]{allen_etal_fig14b.eps}}}}
\caption{(a): Nucleation rate $I$ as a function of shear rate ${\dot{\gamma}}$ for the two dimensional sheared Ising model, computed using DFFS. The nucleation rate peaks at intermediate shear rate. For details of the parameters, please see Ref. \cite{ising_shear1}. (b): Scatter plot of the local density $\rho_x$ ($\rho_y$) of up spins surrounding the largest cluster of up spins in the $x$ ($y$) direction, versus the number of spins $N_c$ in the largest cluster, for configurations in the transition state ensemble, at shear rate $\dot{\gamma}=0.06$. 95\% confidence intervals for the Pearson correlation coefficient $r$ are also shown.  The negative correlation observed for $\rho_x$, but not for $\rho_y$, demonstrates that coalescence along the $x$ direction, driven by shear, plays a role in the nucleation mechanism. [Reproduced with permission from Ref \cite{ising_shear1}].\label{fig:shear1} }
\end{center}
\end{figure}

Nucleation processes under shear are a class of nonequilibrium rare event problems to which  FFS should be able to make a  valuable contribution, since they are scientifically and technologically important and in many cases remain poorly understood. As a test case, we used FFS to study nucleation in a sheared two-dimensional Ising model \cite{ising_shear1,ising_shear2}. The lattice was sheared by periodically randomly selecting a row and shifting this and all higher rows to the right by one lattice site. This imposes a linear ``velocity'' profile. The spins were simulated with Metropolis Monte Carlo dynamics. We used DFFS, with the number of up spins as our order parameter, to compute the rate of nucleation of the stable state with a majority of up spins, starting from the ``metastable'' state with a majority of down spins, in the presence of a weak external field favouring the up state [in the absence of shear, the free energy barrier to nucleation was $\approx 22k_BT$] \cite{barrier}. This system is a poor model for most experimental systems, because particle transport is not modelled, and because the velocity profile is imposed rather than being determined by the system itself. Nevertheless, the computed nucleation rate showed interesting behaviour as a function of shear, with a peak at intermediate shear rate, as shown in Figure \ref{fig:shear1}a.

Understanding the physical mechanisms underlying the nonmonotonic trend in Figure \ref{fig:shear1}a proved to be more difficult than calculating the rate itself. Simply comparing TSE configurations at different shear rates did not explain the effect of shear on the nucleation rate. We therefore resorted to devising modified shear algorithms, to test various hypotheses. For example, an algorithm with randomised shear direction (to eliminate shear-induced cluster breakup) removed the decrease in nucleation rate at high shear rate \cite{ising_shear1}. We also devised an unusual way of analysing the TSE, to test the hypothesis that shear-induced cluster coalescence was important in the enhancement of nucleation at low shear rates. We constructed an order parameter which we expected to be coupled to coalescence: the local density $\rho_x$ of up spins close to the largest cluster in the $x$ direction. Since the shear is applied in the $x$ direction, we postulated that, if coalescence is important, then configurations with large values of $\rho_x$ would have greater tendency to nucleate than those with small $\rho_x$. However, we did not expect such an effect for $\rho_y$. This could be tested by making scatter plots of $\rho_x$ versus the largest cluster size $N_c$, for TSE configurations (which all have the same committor value $P_B=0.5$). If coalescence is important, both $N_c$ and $\rho_x$ should contribute to the committor, so that TSE configurations with large $N_c$ will tend to have small $\rho_x$, and {\em{vice versa}}. We therefore expected negative correlation between $\rho_x$ and $N_c$, but not between $\rho_y$ and $N_c$, in the presence of shear only. This was indeed observed, as shown in Figure \ref{fig:shear1}b, allowing us to conclude that shear-induced cluster coalescence is an important factor, at least in this model. The contrast between the behaviour of this model and Classical Nucleation Theory was discussed in a follow-up work  \cite{ising_shear2}, in which we investigated the effect of the external field strength. 

The example discussed here is highly simplified in comparison to realistic sheared nucleation problems. However, even from this simple example, it is clear that nucleation problems under shear hold much potential, as well as presenting new challenges, for  rare event simulation methods such as FFS.

\section{Challenges and future directions}\label{sec:chall}
The development of rare event simulation methods in general, and FFS in particular, is far from complete, and many challenges remain. Some challenges, relating to computational efficiency, parameter optimisation and the choice of order parameter have already been addressed, resulting in the optimised versions of FFS \cite{borrero2007,borrero2008} discussed in this review. These improved algorithms should  extend the method's applicability to more computationally expensive and/or scientifically complex problems.  It is likely that further fruits will be gained by combining advances made in different areas. For example, the Voronoi tessellation approach (Section \ref{sec:voronoi}) could be combined with committor-based order parameter optimisation schemes (Section \ref{sec:lse}). New FFS variants could also be developed by exploiting the analogy with polymer sampling discussed in Section \ref{sec:variants}. New developments will be driven by new users and new applications and to encourage this we believe that FFS should be implemented as soon as possible in widely used simulation packages.

One important issue is the possible presence of intermediate metastable states between A and B. These are a common feature of protein folding problems, and are likely to occur in many other contexts as well. With current FFS methods, partial paths which get stuck in these intermediate states will prove extremely expensive. This should prove a fruitful area for methodological development, especially if it is possible automatically to detect intermediate metastable states. Such methods could for example be used for ``landscape exploration'' in cases where the location of the B state is not known {\em{a priori}}. This would be useful for a variety of problems including the dynamics of  glasses and protein conformational changes. 

A second, related issue concerns systems with multiple alternative reaction channels. These are problematic for many rare event simulation methods. Dynamic sampling schemes like TPS and TIS, in which  new transition paths are generated from old ones, have difficulty finding previously unexplored reaction channels (although replica exchange approaches can help \cite{bolhuis2008,vanerp2008}). Since FFS is a static scheme in which new paths are generated from scratch, it should in principle be able to  explore all reaction channels. However, the choice of order parameter is likely to be crucial.  For example, Sear \cite{sear2008} demonstrated that DFFS can fail to give the correct rate constant for a nucleation model in a two dimensional landscape where the dynamics evolves much more slowly in one dimension than in the other, if the order parameter involves only the fast coordinate. This is because the system does not fully explore the phase space along the slow coordinate and thus misses important transition paths. Juraszek {\em{et al}}  \cite{jurasczek} also reported that DFFS produced an incorrect rate constant for a protein folding problem where there were several possible folding pathways. This may have been due to undersampling of the ensemble of configurations at the first interface  \cite{jurasczek}; order parameter optimisation might be advantageous for such problems \cite{vega,borrero2008}, although it is not immediately clear whether one can optimise the order parameter to allow sampling of two reaction channels simultaneously. Further investigation of the performance of FFS-like methods for problems with multiple reaction channels is clearly needed, and progress in this direction has recently been initiated in the context of both TPS \cite{jutta} and FFS \cite{borrero2009}.

Another important topic concerns how the method explores path space. In FFS, the segments of a transition path, once laid down, cannot be changed. If the nascent path turns out to be unfavourable, it will fail to reach the B state and the only option is to start again with a new path [for the BG or RB variants]. By contrast, in TPS or TIS, new paths are generated by shooting forwards and backwards in time, so that  the initial segments of a path can relax in phase space. This offers likely advantages in complex free energy landscapes \cite{vanerp2008}. However, TPS and TIS are not suitable for nonequilibrium problems. It may be possible to devise FFS-like methods which sample complex landscapes more efficiently, by more intelligent sampling of the configurations at the interfaces, or by self-consistent determination of the order parameter \cite{borrero2007}.  

Other directions of great interest are the idea  of using time itself as an order parameter \cite{vanerp2}, and the potential for borrowing ideas from related methods in which trajectories are pruned or enriched continuously depending on the value of the Lyapunov exponent \cite{tailleur}. Finally, we note that the above discussion has assumed that the rate of transitions from A to B is constant in time ({\em{i.e.}} that our rare event is a Poisson process). This is not always the case  \cite{visco2008,visco2009}. It remains to be seen whether FFS-like methods can be devised for non-Poissonian rare event problems.

\ack
RJA was funded by the Royal Society of Edinburgh and by the Royal Society. CV was funded by the EPSRC under grant EP/E030173 and by the European Union via a Marie Curie Individual fellowship. PRtW's contribution is part of the research program of the
``Stichting voor Fundamenteel Onderzoek der Materie (FOM)", which is financially supported by the ``Nederlandse organisatie voor
Wetenschappelijk Onderzoek (NWO)''. The authors acknowledge the contributions of Daan Frenkel and Patrick Warren to the development of the FFS method, and helpful discussions with Peter Bolhuis, Christoph Dellago, Marco Morelli, Eduardo Sanz and Sorin T{\u{a}}nase-Nicola. We are grateful to Fernando Escobedo for providing us with a preprint of Ref. \cite{escobedo} and for the use of several figures from his work.

\section*{References}

\bibliography{allen_etal_topicalreview}

\end{document}